# Fluctuation-Driven Morphological Patterning: A Novel Approach to Morphogenesis

Oded Agam[1] and Erez Braun[2]

[1]The Racah Institute of Physics, Edmond J. Safra Campus, The Hebrew University of Jerusalem, Jerusalem 9190401, Israel.

[2] Department of Physics and Network Biology Research Laboratories, Technion-Israel Institute of Technology, Haifa 32000, Israel.

**Abstract**

Recent experimental investigations into *Hydra* regeneration revealed a remarkable phenomenon: the morphological transformation of a tissue fragment from the incipient spherical configuration to a tube-like structure - the hallmark of a mature *Hydra* - has the dynamical characteristics of a first-order phase-transition, with calcium field fluctuations within the tissue playing an essential role. This morphological transition was shown to be generated by activation over an energy barrier within an effective potential that underlies morphogenesis. Inspired by this intriguing insight, we propose a novel mechanism where stochastic fluctuations drive the emergence of morphological patterns. Thus, the inherent fluctuations determine the nature of the dynamics and are not incidental noise in the background of the otherwise deterministic dynamics. Instead, they play an important role as a driving force that defines the attributes of the pattern formation dynamics and the nature of the transition itself. Here, we present a simple model that captures the essence of this novel mechanism for morphological pattern formation. Specifically, we consider a one-dimensional tissue arranged as a closed contour embedded in a two-dimensional space, where the local curvature of the contour is coupled to a non-negative scalar field. An effective temperature parameter regulates the strength of the fluctuations in the system. The tissue exhibits fluctuations near a circular shape at sufficiently low coupling strengths, but as the coupling strength exceeds some critical value, the circular state becomes unstable. The nature of the transition to the new state, namely whether it is a first-order-like or a second-order-like transition, depends on the temperature and the effective cutoff on the wavelength of the spatial variations in the system. It is also found that entropic barriers separate the various metastable states of the system.

## 1. Introduction

Morphogenesis in animal development is a pattern formation dynamical process involving a complex interplay of biochemical, mechanical, and electrical processes. For the biochemical sector, the Turing mechanism [1] and its extensions (e.g., see Refs. [2, 3, 4]) have long been considered the prominent paradigm in this field. According to it, complex spatial patterns of two reacting species, activators, and inhibitors, emerge through a self-organization process. Specifically, when the diffusion rate of the inhibitor significantly exceeds that of the activator, and the interactions between the two species are non-linear, a spatial instability arises that eventually stabilizes in one of a few stable patterns; for a review, see Refs. [5, 6]. The corresponding patterns have been observed in several chemical, biological, and ecological systems [7, 8, 9, 10, 11].

However, the Turing mechanism presents several shortcomings that restrict its direct applicability to morphogenesis in developmental processes. One significant limitation is its focus on biochemical (signaling) processes as the exclusive drivers of tissue patterning, overlooking mechanical processes and the evolution of the underlying tissue geometry. It also disregards the two-way feedback processes between bio-signaling and mechanics, which are responsible for the emergence and stabilization of the body form during morphogenesis. Till now, the interplay between mechanical and biochemical processes remains elusive [12]. Moreover, even when focusing solely on the biochemical processes, self-organized developmental dynamics in certain systems face limitations. Specifically, in Turing-like reaction-diffusion mechanisms, the reliance on an unrealistic separation of scales in the diffusivities of the system's constituents [13] often requires fine-tuning of parameters [14]. Multiple strategies have been proposed to address this problem. One approach involves extending the model to encompass more than two interacting species [15, 16, 17, 18]. A second avenue of exploration entails considering the network topology of Turing systems [19, 20], while a third extension involves adding fluctuations (noise) to the otherwise deterministic model, which can originate from either external sources or from the discrete nature of the reacting entities [21, 22, 14, 23].

Yet, fundamental problems still persist. The first is the lack of robustness, wherein resulting patterns exhibit sensitivity to initial and boundary conditions and/or noise, impeding the formation of intricate patterns, such as those observed in ocular development [24]. The second limitation is the collapse of the mechanism when accounting for the temporal nature of the reactions in biological systems, as reactions are not instantaneous [25, 24].

In pursuit of constructing a robust mechanochemical model capable of providing a comprehensive explanation for the processes underlying pattern formation, an alternative approach to morphogenesis has emerged [26, 27, 28]. This approach centers on the interplay between diffusing chemicals and the tissue's physical structure by introducing couplings of diffusing morphogens with some configurational properties of the tissue, such as its local curvature or stretching deformations [29, 30, 31, 32]. However, in essence, these mechanochemical models are merely extensions of the Turing instability, where one of the tissue's geometrical properties, say the local curvature, assumes the role of the long-range inhibitor field, while a diffusing morphogen serves as a local activator field. In particular, the instability towards pattern formation is, in general, revealed by a linear stability analysis of the uniform stationary state with respect to non-uniform spatial perturbations (although, when dynamics is far from equilibrium, final patterns may be different from those associated with the most unstable modes of the system [33]). However, as we will show below, linear stability analysis may not always offer an adequate framework for understanding the emergence of patterns.

In this work, we propose an alternative mechanism for pattern formation that relies on fluctuations as the main driving force. From our perspective, fluctuations are not merely inescapable background noise. Instead, they play a prominent role in driving the system's dynamics. The proposed mechanism draws inspiration from our observations of the morphological dynamics during whole-body *Hydra* regeneration from a small tissue segment. It suggests that the main morphological transition from the incipient approximately spherical shape to a tube-like structure, characterizing the body-form of a mature *Hydra*, is a phase transition [34]. An intriguing aspect of this transition is its remarkably short duration, approximately 7 minutes. This rapid transformation follows an extended period (days) of no significant morphological changes, and it stands in stark contrast to the considerably longer time scales associated with the underlying mechanical and chemical relaxation times, which are around 100 minutes [35]. The sudden nature of the morphological transition suggests that the dynamics involve activation over a barrier that separates two minima within an effective potential energy landscape.

Each minimum represents a distinct morphological state of the system. One is associated with the nearly spherical shape, while the other corresponds to the elongated state. Thus, unlike the Turing mechanism, our proposed mechanism of morphological patterning cannot be analyzed using linear stability analysis, as the system consistently exhibits local stability. Our recent study [36] corroborates these findings by demonstrating the presence of stochastic morphological swings between two distinct configurations: sphere-like and tube-like configurations. When subjected to external periodic perturbation, these morphological transitions exhibit a resonance-like response as a function of the noise level. Such behavior is consistent with the stochastic dynamics of a system characterized by an underlying double well potential modulated by external forcing.

In identifying the factors governing the *Hydra*'s tissue morphological transition, it is essential to acknowledge that the short duration of the transition precludes significant cell division [37], cell migration [38, 39, 40], or positional rearrangements [41]; these processes take place on time scales of the order of hours or days. Consequently, the transition is primarily driven by the internal active muscle (actomyosin) forces of the tissue and the hydrostatic pressure within the enclosed cavity of the hollow spheroidal tissue [42]. The activation of the muscle contractions is contingent on the concentration of free calcium ions ($Ca^{2+}$), characterized by spatio-temporal fluctuations. Thus, the calcium, which constitutes a scalar field residing on the tissue, plays a prominent role in the pattern formation process [43]. Specifically, during the period preceding and throughout the morphological transition, the *Hydra* tissue displays fluctuating calcium activity [35], localized in space and exhibiting temporal correlations over a typical time scale of approximately an hour. Additionally, the calcium activity exhibits a negative correlation with the local curvature of the tissue, i.e., higher activity is observed in regions where the curvature is low [34].

To show how these elements can navigate the system in the configurational space toward a final stable pattern, we employ a simple one-dimensional model: a closed contour (representing a one-dimensional tissue) embedded in two dimensions. This model incorporates a dynamic field coupled to the contour's curvature and takes into account the influence of hydrostatic pressure through an area constraint on the contour enclosure. Interestingly, this tractable model demonstrates the main aspects of morphological patterning observed in complex biological tissues. A key finding arising from the analysis of this model is the central role played by morphological barriers. These barriers are instrumental in shaping the nature of the phase transition, whether first or second order and determining the available stable morphological states of the system.

The rest of the article unfolds as follows: In the next section, we introduce the model, and in the subsequent one, we perform a linear stability analysis of the circular shape of the contour. This analysis demonstrates that the circular shape remains stable below a certain threshold of the field-curvature coupling strength. Sections 4 and 5 identify configurations representing the system's lower energy and metastable states. Moving on to Section 6, we trace a downhill path from the circular morphology to an elongated shape, highlighting an anomaly in our earlier linear stability analysis. Here, we explain how an effective upper cutoff on the spatial frequencies of the system contributes to re-stabilizing the circular shape through a barrier. Section 7 briefly discusses entropic barriers and Section 8 introduces the dynamic Langevin equations that describe the transition from the circular state to the elongated contour. Section 9 presents the results from Monte Carlo simulations approximating these dynamic equations. Lastly, in Section 10, we discuss the results.

## 2. The Model

To present the basic idea of fluctuation-driven pattern formation, we consider a simple one-dimensional biological tissue embedded in a two-dimensional space. The tissue is represented by a closed curve in the $xy$ plane, expressed parametrically as $\boldsymbol{r}(s)=[x(s),y(s)]$, where $s$ is the arclength parameter, $ds^2=dx^2+dy^2$. In addition, we introduce a non-negative scalar field, $\phi(s)\geq 0$ that resides on the tissue. The energy functional associated with an equilibrium state of the system is chosen to be:

$$E=\oint ds\left[\kappa H^2+\sigma+\beta\left(\frac{\partial\phi}{\partial s}\right)^2+U(\phi)+\eta H^2\phi\right], \tag{2.1}$$

such that the probability density for the various configurations of the system in the equilibrium state is given by:

$$P[\boldsymbol{r}(s),\phi(s)]=\frac{1}{Z}\exp\left(-\frac{E}{T}\right), \tag{2.2}$$

where $T$ is the effective temperature (controlling the magnitude of fluctuations in the system), and $Z$ is the normalization constant.

The first term in the above energy functional is a bending energy term, where $\kappa$ is the bending modulus, and $H=\boldsymbol{r}'(s)\times\boldsymbol{r}''(s)$ is the local line curvature, i.e., the inverse radius of the osculating circle at point $\boldsymbol{r}(s)$ on the contour. Here, the prime denotes derivative with respect to the arclength, and the cross product (in two-dimensional space) is a scalar quantity whose sign determines whether the osculating circle resides inside or outside the region enclosed by the tissue curve ($H>0$ and $H<0$ respectively). The elastic part of the energy does not contain a term linear in the curvature since the integral of the line curvature along a closed contour in the plane is a topological invariant [44]. The second contribution to the energy, represented by the constant $\sigma$, is the line tension energy. The third term in Eq. (2.1) accounts for the field stiffness concerning spatial variations. It captures the resistance of the system to short-range field variations. The fourth term is a potential that characterizes the nature of the $\phi$ field fluctuations. In a regenerating *Hydra*'s tissue, this potential takes the form of a tilted double-well potential that accounts for the excitable nature of calcium fluctuations within the tissue [35]. Finally, the last term in Eq. (2.1) represents the coupling between the curvature and the scalar field. For $\eta>0$ (which will be assumed hereinafter), this coupling encourages high value of $\phi$ in regions of low curvature and conversely promotes low values in regions of high curvature. In essence, this coupling establishes a correlation between the curve's line curvature and the field's spatial variations, influencing the system's overall pattern formation and dynamics.

In what follows, we shall further simplify the model by setting $\beta=0$ demonstrating that, unlike the Turing mechanism, diffusion does not play a role here. We also choose

$$U(\phi)=\alpha(\phi-\phi_0)^2;\quad \phi>0, \tag{2.3}$$

where $\phi_0$ is a constant that can be set to unity without loss of generality. Additionally, it is possible to set one of the parameters, $\alpha$, $\kappa$, $\sigma$ or $\eta$, to unity. This choice can be accommodated by absorbing this parameter into the temperature $T$, along with proper redefinitions of the other parameters. In what follows, we set $\alpha=1$. Finally, to reduce the number of parameters and maintain a dimensionless model, we choose to work in a system with $\sigma = \kappa$ and where the initial state of the contour is a circle of unit radius. Adhering to these selections, the expression for the energy (1) simplifies to:

$$E = \oint ds\left[\kappa\left(H-1\right)^2 + \left(\phi - 1\right)^2 + \eta H^2 \phi\right] - 2\pi\eta\left(1 - \frac{\eta}{4}\right). \tag{2.4}$$

The last term in the right-hand side of the above equation represents an energy shift by a constant value, which, for convenience, sets the minimal energy of the circular state to zero. In the context of the *Hydra*, an additional force exerted on the tissue stems from the hydrostatic pressure within the hollow region enclosed by the tissue. Due to the internal fluid incompressibility, changes in the volume of this region during the rapid morphological transition are minimal. Thus, one can effectively address the impact of pressure by imposing a fixed-volume constraint. In our simplified model, we adopt an analogous condition by constraining the area enclosed by the contour. Specifically, we maintain this area equal to the area of a unit-radius circle that represents the initial state of the system:

$$A = \frac{1}{2}\oint ds\left|\boldsymbol{r}\left(s\right)\right|^2 = \pi\ . \tag{2.5}$$

## 3. Linear stability analysis of the circular state

The minimal energy of the circular state of the system is reached when $H = 1$ and $\phi = \phi_c = 1 - \eta/2$, where hereinafter we assume $0 < \eta < 2$. In representing small variations of the contour from the circular shape, it is convenient to use a Fourier expansion for the polar representation of the curve,

$$R\left(\theta\right) = R_0\left[1 + \sum_{k=2}^{\infty} a_k \cos\left(k\theta\right) + b_k \sin\left(k\theta\right)\right], \tag{3.1}$$

such that $\boldsymbol{r} = R\left(\theta\right)\left(\cos\theta, \sin\theta\right)$ is a point on the tissue contour, and $0 \le \theta < 2\pi$. Notice that the term $k = 1$ produces a constant shift of the contour. Hence, it is a zero mode that should be removed from (3.1). The area enclosed by the contour is:

$$A = \frac{1}{2}\int_0^{2\pi} d\theta R^2\left(\theta\right) = \pi R_0^2\left[1 + \frac{1}{2}\sum_{k=2}^{\infty}\left(a_k^2 + b_k^2\right)\right], \tag{3.2}$$

and it is fixed to $\pi$ (see Eq. (2.5)) by setting

$$R_0 = \left[1 + \frac{1}{2}\sum_{k=2}^{\infty}\left(a_k^2 + b_k^2\right)\right]^{-1/2}. \tag{3.3}$$

Defining the vectors

$$\boldsymbol{t} = \frac{\partial \boldsymbol{r}}{\partial \theta}, \quad \boldsymbol{t}' = \frac{\partial^2 \boldsymbol{r}}{\partial \theta^2}, \tag{3.4}$$

an infinitesimal element of the arclength is given by $ds = \sqrt{\boldsymbol{t}\cdot\boldsymbol{t}}d\theta$ while the line curvature is

$$H = \frac{\boldsymbol{t}\times\boldsymbol{t}'}{\left(\boldsymbol{t}\cdot\boldsymbol{t}\right)^{3/2}}. \tag{3.5}$$

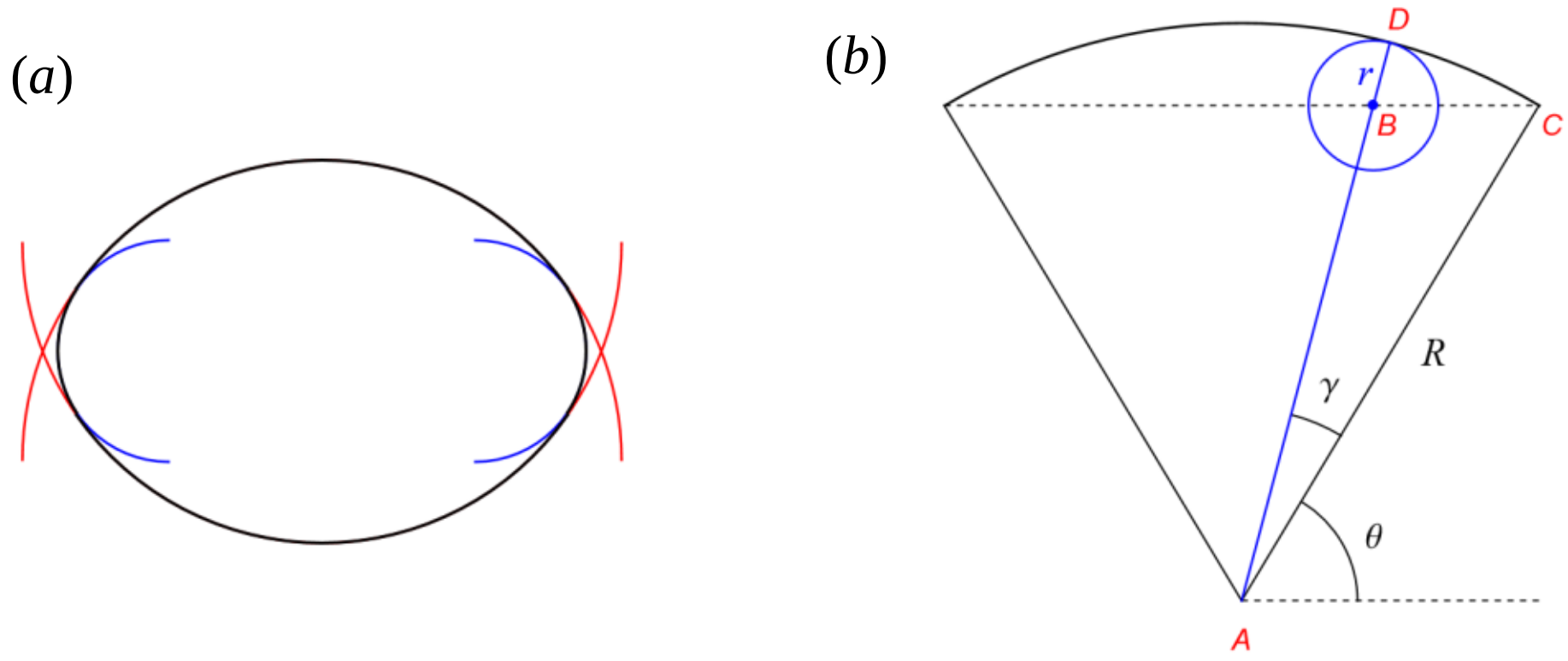

**Figure 1. A two-fold contour shape and its parametrization.** Left: A two-fold symmetric contour made of arcs with two distinct radii, $R$ and $r$. Right: Parametrization of the contour by two angles, $\theta$ and $\gamma$ (with $\theta,\gamma \geq 0$ and $\theta+\gamma < \pi/2$ ), that uniquely define its shape, given that the area enclosed by the contour is maintained $\pi$.

We also expand the scalar field in a Fourier series around the value that minimizes the energy of a circular contour:

$$\phi = 1 - \frac{\eta}{2} + \sum_{k=1}^{\infty}\left[c_k \cos(k\theta) + d_k \sin(k\theta)\right] . \tag{3.6}$$

Expanding the energy (2.4) up to quadratic order in the Fourier coefficients, $a_k$, $b_k$, $c_k$ and $d_k$, it takes the form:

$$\begin{aligned} E = {} & \pi\kappa\sum_{k=2}^{\infty}\left[a_k^2\left(k^2-1\right)^2 + b_k^2\left(k^2-1\right)^2\right] + \pi\sum_{k=1}^{\infty}\left(c_k^2+d_k^2\right) - 2\pi\eta\sum_{k=2}^{\infty}\left(k^2-1\right)\left(a_k c_k + b_k d_k\right) \\ & + \pi\eta\sum_{k=2}^{\infty}\left(k^2-1\right)\left[\frac{2k^2-3}{2} - \frac{4k^2-7}{8}\eta\right]\left(a_k^2+b_k^2\right) . \end{aligned} \tag{3.7}$$

Equation (3.7) is a sum over quadratic forms, each associated with a distinct value of $k$. The matrix describing the quadratic form for wavenumber $k$ is:

$$M_k = \pi\begin{pmatrix} \kappa\left(k^2-1\right)^2 + \eta\left(k^2-1\right)\left[\frac{2k^2-3}{2} - \frac{4k^2-7}{8}\eta\right] & -\eta\left(k^2-1\right) \\ -\eta\left(k^2-1\right) & 1 \end{pmatrix} \tag{3.8}$$

Diagonalization of this matrix yields two eigenvalues. One eigenvalue, $\lambda_1$ is positive for any strength of the coupling parameter $\eta$, whereas the second eigenvalue, $\lambda_2$, can become negative. In the limit $k \to \infty$, the asymptotic value of $\lambda_2$ is $\lambda_\infty = 3 - 4/(\eta - 2)$. It becomes negative when $\eta > 2/3$, signifying the instability of the tissue's circular state via short wavelength deformations. This condition holds true for $\kappa > 1/6$. For $\kappa < 1/6$ a slightly lower value of $\eta$ can induce instability. Here $\lambda_2$, as a function of $k$, initially decreases to a negative value and subsequently increases towards the asymptotic value $\lambda_\infty$ which is positive for $\eta < 2/3$. The minimum of $\lambda_2$ is reached for $k = k_{\min}$ where $k_{\min} = \sqrt{\left(8\kappa + 15\eta^2 - 12\eta\right)/\left(12\eta^2 - 8\eta\right)}$. Requiring that this wavenumber is equal to or greater than the minimum possible wavenumber in the sum (3.7), i.e. $k_{\min} \geq 2$, we obtain that the circular state becomes unstable when $\eta > (2/33)\left(5+\sqrt{25+66\kappa}\right)$. From these considerations, it can be concluded that the threshold value of the coupling parameter indicating instability is:

$$\eta_{\mathrm{th}} = \min\left[\frac{2}{33}\left(5+\sqrt{25+66\kappa}\right), \frac{2}{3}\right]. \tag{3.9}$$

It is worth noting that when $\kappa < 1/6$ and $\eta_{\mathrm{th}} < \eta < 2/3$, the instability emerges through the long-wavelength modes of the system. However, even as $\kappa \to 0$, this behavior manifests within a very narrow range of the coupling parameter values, $20/33 < \eta < 2/3$.

## 4. The stable state of the system for a moderate coupling strength

The energy (2.4) reaches its minimum when both $H$ and $\phi$ take piecewise constant values. Hence, the configuration that minimizes the energy consists of circular arcs joined together to form a closed contour. Specifically, a contour composed of two distinct types of arcs, exemplifying a two-fold symmetry, has the structure shown in Fig. 1a. We will demonstrate below that this particular morphology becomes energetically favorable when the coupling parameter $\eta$ falls within a moderate range below the threshold for instability (3.9), $\eta_c < \eta < \eta_{\mathrm{th}}$ ($\eta_c$ will be defined later; see Eq. (6.1)).

It is convenient to parametrize the contour using the two angels, $\theta$ and $\gamma$, defined in Fig. 1b. Applying the sine theorem to the triangle ABC, we can relate the two radii of the circular sections, $r$ and $R$:

$$r = R\left(1-\frac{\sin\theta}{\sin(\theta+\gamma)}\right) \quad \text{with } \theta,\gamma \ge 0 \text{ and } \gamma+\theta < \frac{\pi}{2}. \tag{4.1}$$

Then, by simple trigonometrical considerations, the area enclosed by the contour is:

$$A = R^2\left\{\pi - \sin 2\theta + \frac{2\sin\theta}{\sin(\theta+\gamma)}\left[-1+\frac{2}{\sin(\theta+\gamma)}\left[\sin\gamma - (\theta+\gamma)\left(2-\frac{\sin\theta}{\sin(\theta+\gamma)}\right)\right]\right]\right\}. \tag{4.2}$$

Imposing the area constraint (2.5) determines the value of $R$. The energy (2.4) for the two-fold symmetric morphology then reduces to:

$$\begin{aligned} E_2(\theta,\gamma) = {} & 4\left(\frac{\pi}{2}-\theta-\gamma\right)R\left[\kappa\left(\frac{1}{R}-1\right)^2 + (\phi_1-1)^2 + \frac{\eta}{R^2}\phi_1\right] \\ & + 4(\theta+\gamma)r\left[\kappa\left(\frac{1}{r}-1\right)^2 + (\phi_2-1)^2 + \frac{\eta}{r^2}\phi_2\right] - 2\pi\eta\left(1-\frac{\eta}{4}\right), \end{aligned} \tag{4.3}$$

where

$$\phi_1 = \max\left(0, 1-\frac{\eta}{2R^2}\right) \quad \text{and} \quad \phi_2 = \max\left(0, 1-\frac{\eta}{2r^2}\right) \tag{4.4}$$

represent the optimal field values that minimize the energy on each circular arc. The energy is now minimized over the parameters $\theta$ and $\gamma$, resulting in the minimizing angles, $\theta_{\min}$ and $\gamma_{\min}$, which determine the contour's shape. Several examples of contours corresponding to this minimal energy are illustrated in Fig. 2, for a fixed coupling parameter, $\eta = 0.5$, and various values of the bending modulus $\kappa$. Within each shape in Fig. 2, we marked the minimal energy, $E_{\min}^{(2)} = E_2(\theta_{\min}, \gamma_{\min})$ along with the optimal values of $\theta$ and $\gamma$. A negative value of $E_{\min}$ implies that the corresponding configuration is lower than that of the circular state (whose energy is zero). As explained below, these configurations represent the stable morphologies of the tissue.

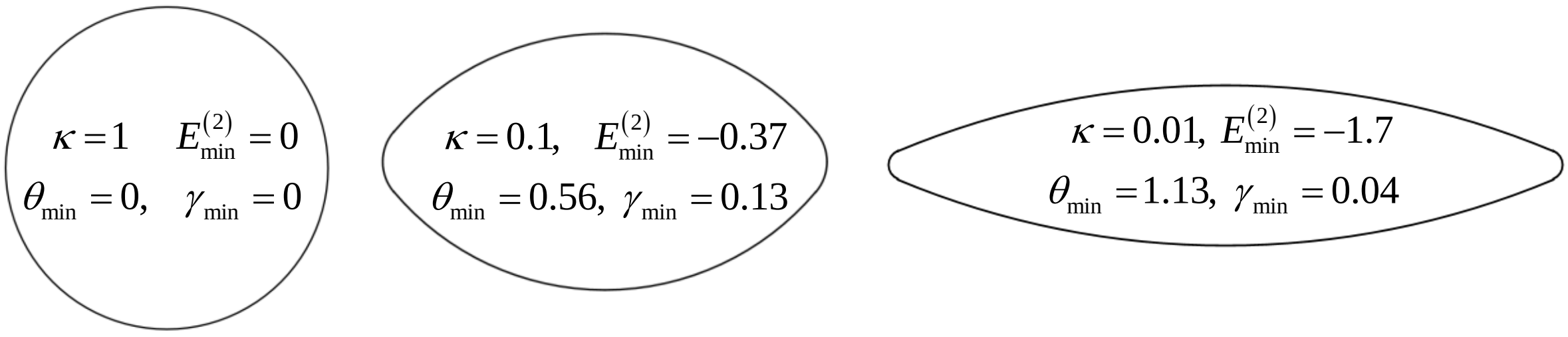

**Figure 2. Examples of two-fold symmetric contour shapes.** The contour shapes are obtained for a fixed value of the parameter, $\eta = 0.5$, and various values of the dimensionless bending modulus $\kappa$. Within each contour the angles $\theta_{\min}$ and $\gamma_{\min}$ that minimize the energy (4.3) are specified. $E_{\min}^{(2)}$ is the energy of the corresponding contour. Negative values of this energy imply that the circular configuration (whose energy is zero) is either unstable or metastable.

The examples depicted in Fig. 2 demonstrate that despite $\eta = 0.5$ being well below the critical value for a circle linear instability (3.9), their energies are negative, i.e., the energy of the two-fold symmetric morphologies is lower than that of the circular state. This finding indicates that while the circular shape of the tissue may be locally stable under certain conditions, a different configuration with lower energy is available. The system can make a transition to this state in the presence of fluctuations. This stable two-fold symmetric shape (demonstrated in Fig. 2 for various system parameters) is not captured by the naïve linear stability analysis previously discussed.

As a final comment, we add that the minimization of energy for a non-symmetric contour shape, such as varying the radii for the left and right blue arcs in Fig. 1a independently, results in energies exceeding that of the symmetric configuration depicted in Fig. 2. Furthermore, as we show in the following section, contour configurations with higher-fold symmetries possess higher energy levels when the coupling parameter assumes moderate values (as will be specified below). Thus, the elongated morphology displayed in Fig. 2 emerges as the lowest energy configuration of the system.

## 5. Metastable morphologies

It is instructive to examine alternative configurations characterized by local minima of energy. In this section, we discuss contour shapes manifesting an *n*-fold symmetry. Fig. 3 portrays illustrative configurations of such contours (for the same parameters as in the middle panel of Figure 2, i.e., and $\eta = 0.5$).

The energy for *n*-fold symmetric contours constructed from arcs with two distinct radii is derived by a direct extension of the calculation expounded in the preceding section. The details of this calculation are presented in Appendix A. The characterization of an *n*-fold symmetric contour necessitates the specification of two parameters, which can be conveniently chosen as angles, denoted as $\theta$ and $\gamma$, akin to the depiction presented in Fig. 1 (see Fig. A1 in Appendix. A). The minimization of energy, denoted as $E_n\left(\theta,\gamma\right)$ over these two angles, is carried out numerically. This energy as a function of $n$, $E_{\min}^{(n)} = E_n\left(\theta_{\min},\gamma_{\min}\right)$, is presented in Fig. 4a. For clarity, it is normalized by the absolute value of the energy associated with the 2-fold symmetric contour, $\left|E_{\min}^{(2)}\right|$. The calculation is performed for the same parameters as in Fig. 2, $\kappa = 0.1$ and $\eta = 0.5$. This figure shows that the elongated two-fold symmetric configuration is the global minimum of the energy. However, it also unveils that the energy of the three-fold configuration deviates by a modest margin of approximately 7% from the two-fold symmetric state. This observation signifies that, under suitable conditions (i.e., low temperatures, see below), the system may transiently adopt this metastable three-fold state and persist in it for a considerable duration. A similar trend holds true for other *n*-fold symmetric contours, as their minimal energy lies within 10% compared to that of the two-fold symmetric state.

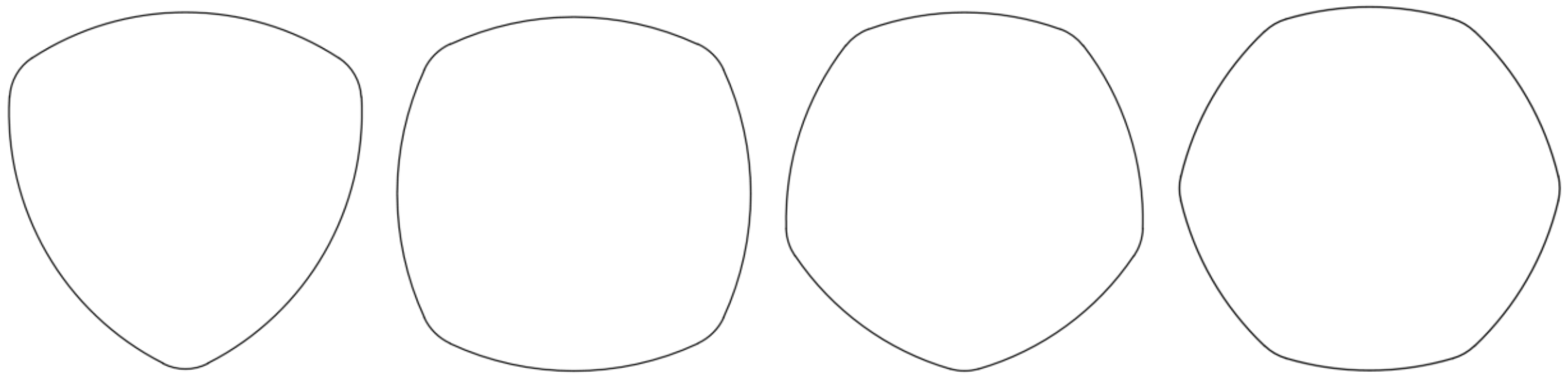

**Figure 3. Examples of *n*-fold symmetric contour shapes.** The configurations are obtained by minimizing the energy (2.4), assuming the contour is *n*-fold symmetric, $\kappa = 0.1$ and $\eta = 0.5$.

The two-fold symmetric state ceases to be the global energy minimum when the coupling strength attains a sufficiently elevated level. This assertion is demonstrated by plotting the normalized difference between the three-fold and the two-fold symmetric configurations, $\left(E_{\min}^{(2)} - E_{\min}^{(3)}\right)\big/E_{\min}^{(2)}$, as a function of the coupling parameter, for various bending modulus values. The outcomes depicted in Fig. 4b indicate that the normalized difference becomes negative for a coupling parameter exceeding approximately one (the precise value depends on $\kappa$), signifying a lower energy minimum of the three-fold symmetric state. Yet, as the coupling strength crosses this threshold, the energy of higher symmetric configurations becomes even lower, although the energy difference between these configurations is minute. For specific examples of this behavior, see the tables at the end of Appendix A. The linear stability analysis of these *n*-fold symmetric shapes, discussed in Appendix B, shows that these shapes are locally stable. Namely, they represent metastable states.

It is important to recognize that the energy value and the stable nature of the $n$-fold symmetric shapes alone do not exclusively govern the selected morphology towards which the system relaxes from the circular state. Additional determinants encompass the nature of the attraction basin for a given state. In cases of significant fluctuations (which is the focus of this article), the phase space volume available due to fluctuations in the proximity to the minimum plays a significant role. Indeed, in Appendix B we show that renormalization of the energies of the *n*-fold symmetric configurations by fluctuations stabilizes the 2-fold symmetric state when the temperature controlling the level of fluctuations is high enough.

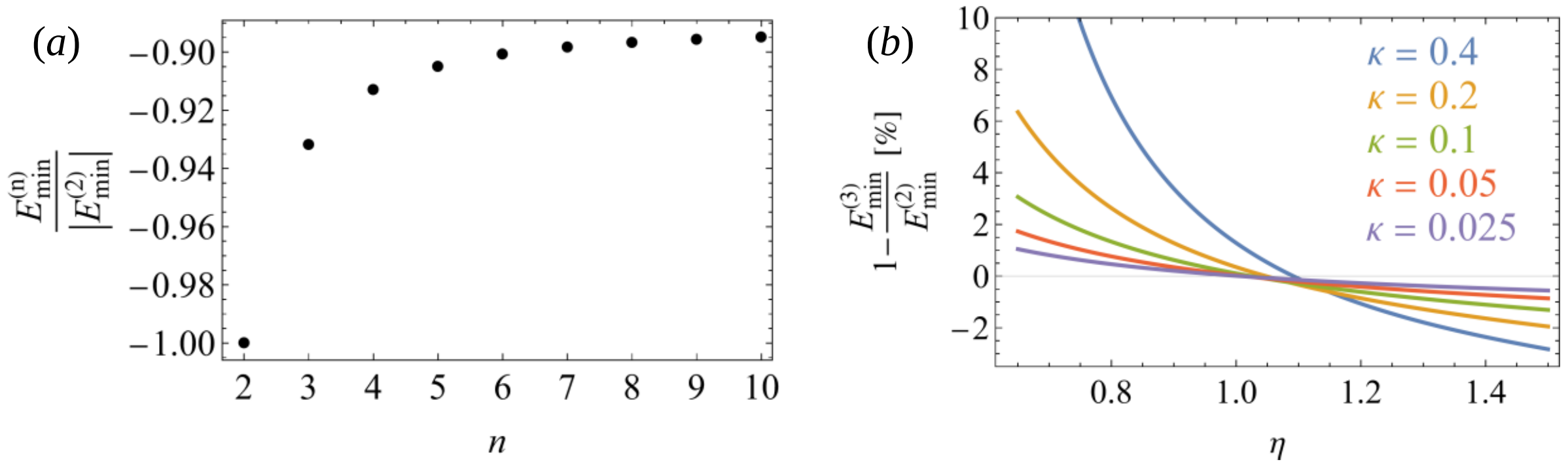

**Figure 4. The energy of metastable states.** (*a*) The ratio of the energy of an *n*-fold symmetric shape to that of the two-fold symmetric state as a function of $n$ for $\kappa = 0.1$ and $\eta = 0.5$ (the same parameters used in Fig. 3). (*b*) The normalized energy difference between the 3-fold symmetric configuration and the two-fold symmetric state as a function of the coupling parameter, for various values of the dimensionless bending modulus. The change of sign around $\eta \simeq 1.1$ indicates that when the coupling is strong enough, high symmetry configurations are the energetically favorable states of the system.

## 6. First and second-order transitions

Based on the above results, it becomes apparent that the circular and elongated configurations depicted in Fig. 2 represent two distinct system minima. Thus, one may expect a transition between these states via activation over the barrier separating them. Yet, as we argue below, there is a downhill path in the contour configuration space, allowing a direct transition of the system from the circular state to the elongated state without crossing a barrier. To demonstrate that, in Fig. 5, we plot the energy, $E_2(\theta,\gamma)$ as a function of $\theta$ for a set of incrementing values of $\gamma$. Here, the bending modulus is $\kappa = 0.1$, while the coupling parameter, $\eta = 0.35$, is well below the threshold value given by Eq. (3.9). This figure shows that, for a fixed value of $\gamma$, the energy $E_2(\theta,\gamma)$ displays a minimum at a finite value $\theta$, and as $\gamma$ increases the depth of this minimum increases, ultimately leading to a global minimum at some finite value of $\gamma = \gamma_{\min}$. The continuous change of these minima, as $\gamma$ increases from zero to $\gamma_{\min}$ defines a continuous set of pairs $(\theta,\gamma)$. These pairs specify a trajectory within the contour's configuration space, transitioning from the circular state to the stable elongated configuration. Along this trajectory, the energy value progressively decreases at each successive state, indicating a smooth transition between the two shapes without encountering any energy barrier.

Fig. 6 demonstrates the evolution of the instability from the circular state. Each curve in this figure displays part of the closed contour associated with a local minimum of the functional $E_2(\theta,\gamma)$, where $\gamma$ is held fixed, and minimization is with respect to $\theta$. The values of $\gamma$ range between zero and $\gamma_{\min}$. The red and the blue segments indicate circular sections of different curvatures. The observed behavior manifests the localized nature of the instability. Hence, multiple instabilities may develop concurrently along the contour. The competitive dynamics and mutual interactions among these instabilities ultimately dictate the selection of the stable or metastable state towards which the system evolves.

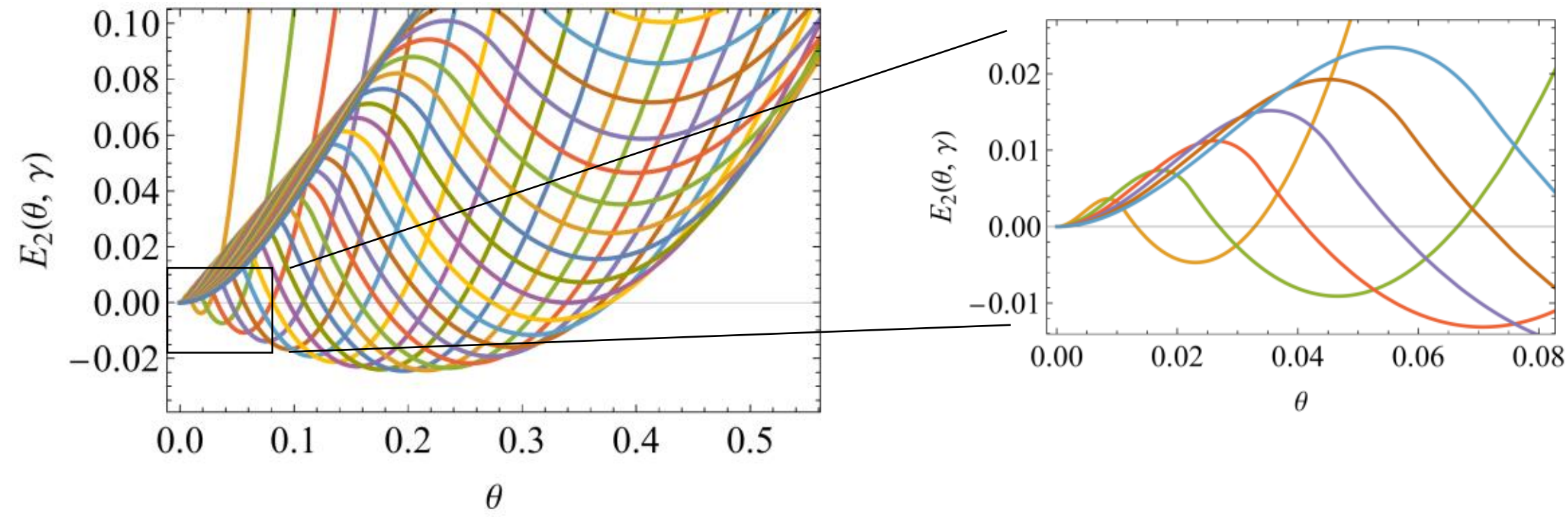

**Figure 5. Limits in the linear stability analysis.** Left: The energy $E_2(\theta, \gamma)$ as a function of $\theta$ for increasing values of $\gamma$ (different colors, with increasing values for the curves from left to right). The envelope of the curves shows that there is a downhill trajectory that takes the system from the circular state into the elongated state. Right: A magnified part of the left panel, showing that for any finite value of $\gamma$ there is a region where the energy curve as a function of $\theta$ first increases before turning down. Thus, for linear stability analysis, the order of the limits $\theta \to 0$ and $\gamma \to 0$ do not commute. The above calculation is performed for $\kappa = 0.1$ and $\eta = 0.35$ where naïve linear stability analysis predicts the system is stable.

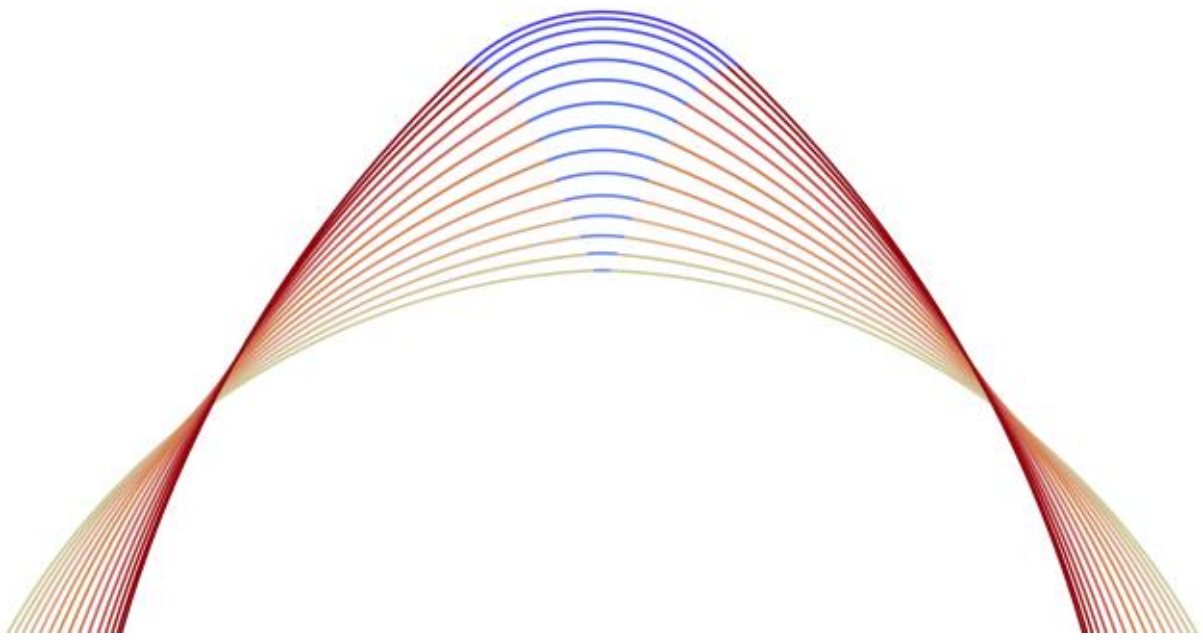

**Figure 6. The evolution of an instability from the circular initial state**. The various curves in this plot show an enlarged view of a section of the contour where the instability emerges from the initial circular state. The colored segments represent circular sections of different curvatures and different values of the scalar field, $\phi$. Here we use a thermometer color code, where blue and red colors denote low and high values of the field, respectively.

As the coupling parameter $\eta$ decreases, the values of $\theta$ and $\gamma$ associated with the global energy minimum continuously diminish towards zero. In other words, the contour associated with the minimal energy becomes closer to a circle. Below some critical value,

$$\eta_c = \sqrt{\kappa} \,, \tag{6.1}$$

the circular configuration becomes globally stable. Formula (6.1) is valid as long as $\eta_c$ is smaller than $\eta_{\text{th}}$, i.e. for $0 < \kappa < 4/9 \simeq 0.44$. For the derivation of (6.1), see Appendix C. The above observations imply that the system undergoes a second-order phase transition as $\eta$ increases beyond $\eta_c$. In other words, one can define a parameter that characterizes the shape of the contour (see Eq. (9.6) below) serves as the order parameter of the system, which is zero for $\eta \le \eta_c$, and increases (from zero) continuously with the coupling parameter as $\eta > \eta_c$.

The apparent contradiction between this scenario and the stability analysis of the circular state, which indicates a first-order transition, can be attributed to an anomaly arising from the vanishing validity range of the linear stability analysis. Indeed, a demonstration of this anomaly is provided by the magnifying view of $E_2(\theta,\gamma)$ in the region $\theta,\gamma \to 0$ depicted in the right panel of Fig. 5. It demonstrates that for any finite value of $\gamma$, the energy first increases as a function of $\theta$ before turning down, apparently implying local stability of the circular state. However, as $\gamma$ decreases, the range of $\theta$ showing this increase in energy becomes smaller and the energy barrier separating the circular state from the global non-circular minimum diminishes. This behavior suggests that the stability analysis (3.8) holds on a vanishingly small range of the parameters $a_k$ and $b_k$

From the above considerations, it follows that as $\eta$ gradually increases from a value below the critical threshold to one above it, the system undergoes a second-order-like phase transition from the circular state to a broken-symmetry state. However, in general, there exists an upper cutoff on the Fourier components of the contour shape, which sets a lower bound on the possible values of $\gamma$ - the parameter that determines the small radius of curvature given Eq. (4.1). The presence of such a cutoff effectively converts the second-order transition into a weak first-order type. It can be incorporated into the model in various ways, for instance, by discretizing the curve representation or any other mechanism that limits fast curvature variations.

## 7. Entropic Barriers

The preceding discussion is framed within the saddle-point approximation, where fluctuation effects are disregarded (mean field approximation). However, given the high-dimensional nature of the configurational space describing the contour shape, one anticipates that fluctuations play an important role by introducing entropic barriers. Specifically, fluctuations force the system to diffusely explore the large configurational space, searching for the optimal trajectory that enables it to undergo a morphological transition. The many possible trajectories taking the system away from the optimal trajectory can be interpreted as entropic barriers that hinder the direct transition along the preferred pathway.

The appearance of entropic barriers, reminiscent of the Levinthal paradox in the context of protein folding [45, 46], can be illustrated with a simple conceptual model [47]: Let $x$ represents a "reaction coordinate" in the high dimensional configurational space of a system, and assume that in the absence of fluctuations the system follows overdamped dynamics in a smooth and attractive potential $V(x)$, with its minimum at $x=0$. In the presence of fluctuations, the system's probability to occupy a state, characterized by $x$ is proportional to $g(x,T)\exp(-V(x)/T)$, where $T$ is the temperature characterizing the fluctuations, and $g(x,T)$ is the fugacity, i.e. $g(x,T)dx$ is the number of different microscopic states for reaction parameter within the range $(x, x+dx)$. For instance, if the configurational space of the system is considered to be a hyper-sphere in $d \gg 1$ dimensions, then $g(x,T)=(2\pi)^{d/2}x^{d-1}/\Gamma(d/2)$, where $\Gamma(z)$ is the Gamma function. The presence of the fugacity factor implies that the effective potential governing the system is $V_{eff}(x)=V(x)-TS(x)$, where $S(x)=\ln g(x,T)$ represents the entropy (in the case of the hypersphere, $S(x)=(d-1)\ln x$). This entropy term introduces an effective barrier in the reaction space, as illustrated in Fig. 7.

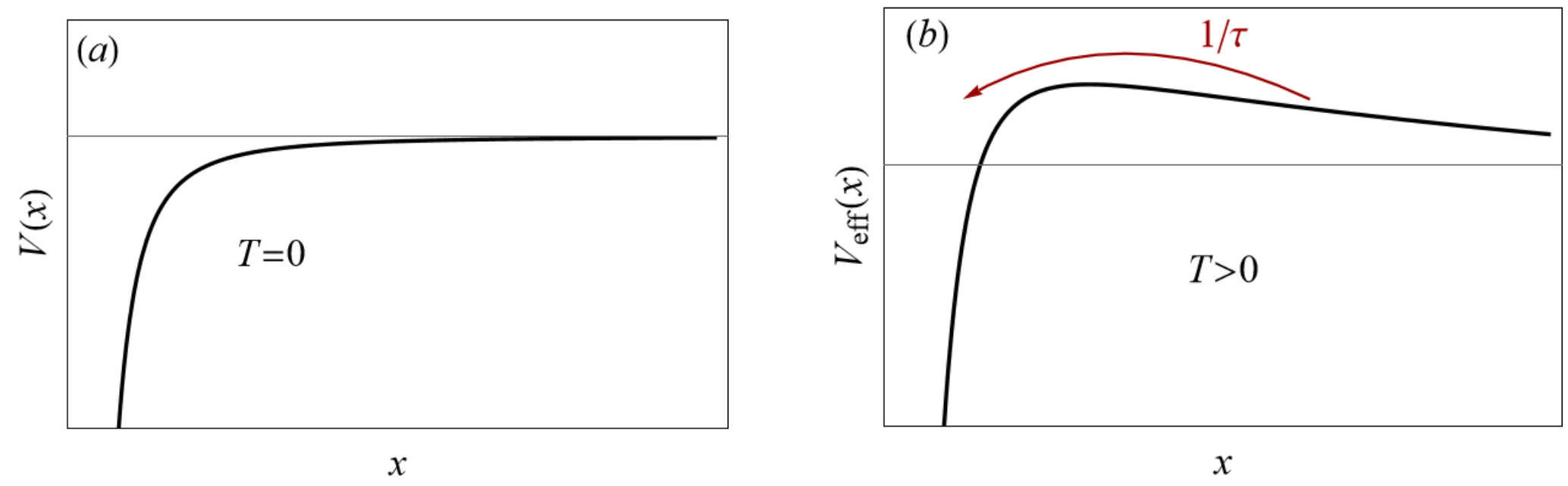

**Figure 7. An illustration of the formation of an entropic barrier by fluctuations**. (*a*) The potential energy, $V(x)$, of an overdamped particle in the absence of fluctuations, i.e., when $T=0$. (*b*) The particle's effective potential, $V_{eff}(x)=V(x)-TS(x)$, when taking into account the entropic barrier due to fluctuations.

Entropic barriers may switch a second-order transition into a first-order one. Hence, fluctuations in our system have two opposing effects. On the one hand, they increase the configurational phase space volume available to the system, impeding a direct transition by creating entropic barriers. On the other hand, fluctuations also serve as the drivers for activation, enabling the system to cross the energy barrier and facilitate the transition. This activation effect becomes particularly relevant when the effective potential features a barrier due to ultraviolet cutoff or other constraints. The balance between the height of the entropic barriers and the probability for activation over these barriers, determined by the spectrum of fluctuations, dictates the nature of the transition and the system's dynamics.

## 8. Equations of motion

Thus far, our focus has been solely on the "equilibrium" or metastable states of the system, which are determined by the probability distribution function (2.2) determined by the energy (2.4). The existence of a multitude of diverse dynamical processes that can converge to the same equilibrium state is well-known. Consequently, inferring unique equations governing the time evolution of a system, from the circular state to the low energy configuration of the system, solely based on its equilibrium states is not feasible. However, assuming dissipation is dominant over inertial effects, the Langevin equation emerges as a natural and generic choice for describing the dynamics of such systems. This stochastic differential equation is widely employed to model systems subject to random fluctuations and dissipative forces. In the context of our dimensionless variables, we consider the simplest form of the (coupled) Langevin equations:

$$\begin{aligned}\frac{\partial\phi}{\partial t}&=f_\phi+\zeta_\varphi(s,t)\\ \frac{\partial\mathbf{r}}{\partial t}&=\boldsymbol{f}_r+\boldsymbol{\zeta}_r(s,t)\end{aligned}\quad,\qquad\text{with the forces given by}\qquad \begin{aligned}f_\phi&=-\frac{\delta E}{\delta\phi}\\ \boldsymbol{f}_r&=-\frac{\delta E}{\delta\mathbf{r}}\end{aligned}\;. \tag{8.1}$$

Here, $s$ parametrizes the contour, $\zeta_\varphi(s,t)$ and the two components of $\boldsymbol{\zeta}_r(s,t)$ represent uncorrelated stochastic noise terms with zero means and zero range correlations, i.e.,

$$\begin{aligned} \left\langle \zeta_\varphi(s,t)\zeta_\varphi(s',t')\right\rangle &= 2T\delta(t-t')\delta(s-s') \\ \left\langle \zeta_r^{(i)}(s,t)\zeta_r^{(j)}(s',t')\right\rangle &= 2T\delta(t-t')\delta(s-s')\delta_{ij} \end{aligned} \tag{8.2}$$

The equilibrium state, described by the probability distribution function (2.2), corresponds to a situation where the system reaches a balance between thermal fluctuations and dissipation. The Langevin dynamics may explore multiple energy minima, leading to a scenario where the system can be trapped in a metastable state for a considerable period before moving to the global equilibrium state. This behavior is particularly relevant in our system, where the energy landscape is expected to contain multiple local minima separated by energy barriers.

In Appendix D, we derive the expressions for the forces in Eq. (8.1). Specifically, the force associated with the dynamics of the $\phi$ field is

$$f_\phi = 2(1-\phi) - \eta H^2 . \tag{8.3}$$

This force is attenuated within regions of pronounced curvature, signifying the scalar field's responsiveness to the system's configuration. This type of interaction between the tissue's local curvature and the scalar field is aligned with our experimental findings in *Hydra* regeneration [34]. The precise biological underpinning of this phenomenon remains elusive; nonetheless, it is reasonable to anticipate that robust regulation of morphogenetic alterations necessitates reciprocal relations between the scalar field and the local tissue’s shape.

The scalar field's impact on the tissue’s morphology is manifested through the force acting on the tissue. Assuming the contour is parametrized by the arclength $s$, the force that acts on an infinitesimal contour segment, at point $\boldsymbol{r}(s)$, in a direction perpendicular to the contour is

$$f_\perp = \kappa\left(2\frac{\partial^2 H}{\partial s^2} + H^3 - H\right) - H(\phi-1)^2 + \eta\left[2\frac{\partial^2 H\phi}{\partial s^2} + H^3\phi\right] + p , \tag{8.4}$$

where $p$ is the pressure that serves as a Lagrange multiplier enforcing the fixed area constraint (2.5). Finally, the force in a direction tangential to the contour is given by:

$$f_\parallel = -2\kappa\frac{dH^2}{ds} + \frac{d}{ds}(\phi-1)^2 - \eta\left(2H\frac{\partial H\phi}{\partial s} + \frac{dH^2\phi}{ds}\right). \tag{8.5}$$

It should be noted, however, that the equation of motion in the tangential direction merely represents a re-parameterization of the contour as it preserves its form. Yet, the discretization of the contour into small segments will result in displacements of these segments along the contour, thus changing their local density.

## 9. Monte Carlo Simulations

The Langevin equations (8.1) are nonlinear stochastic partial differential equations that pose challenges in their numerical solution. Instead, we adopt Metropolis Monte Carlo simulations as an alternative approach to study the system's dynamics. The Metropolis Monte Carlo technique is a numerical method that essentially solves the Fokker-Planck equations arising from the Langevin equations. Although conventionally used for equilibrium property studies, this technique extends its applicability to the investigation of dynamical properties [48]. The idea is that over sufficiently short time intervals, the noise term in the Langevin equation dominates and induces a local random change in the variables $\phi$ and $\boldsymbol{r}$. This change serves as the Monte Carlo step trial change. The decision to accept or reject this change follows the standard Metropolis criterion, which tends to steer the system toward lower energy regions. As a result, the Monte Carlo dynamics with local changes emulates the actual dynamics of the system, over time scales significantly longer than the Monte Carlo time step and substantially shorter than the "microscopic" characteristic times of the system (in units of the Monte Carlo time steps which do not directly correlate with physical time). This allows us to explore the dynamic behavior of the system in order to corroborate the physical picture presented above.

The specific procedure employed for Monte Carlo calculations is detailed in Appendix E. It is instructive to introduce a parameter that characterizes the contour's shape to present the outcome of these simulations. This parameter, akin to the order parameter in the context of phase transitions, can be defined as follows:

$$\Lambda = 1 - \frac{4\pi A}{P^2} \tag{9.6}$$

where $A = \pi$ is the area enclosed by the contour and $P$ denotes its perimeter [34]. For a circular contour, this shape parameter is zero. However, for any deviation from the circular state, the parameter becomes positive, and as the aspect ratio of the elongated shape of the contour approaches infinity, $\Lambda \to 1$.

We begin by showcasing the outcomes of the Monte Carlo for $\kappa = 0.1$ and $\eta = 0.5$ , a value well below the threshold coupling (3.9) calculated from linear stability analysis ($\eta_{\mathrm{th}} = 0.6437...$). The short wavelength cutoff is set to be $P/30$, and the temperature is $T = 10^{-3}$. Panels a & b of Fig. 8 depict the system's evolution through selected examples of the contour shape at different time points, accompanied by the time-dependence of the shape parameter, $\Lambda(t)$. The colors of the contour represent the intensity of the field $\phi$, using a thermometer color code (Mathematica®), where blue and red signify low and high values of the field, respectively. The shape parameter remains approximately zero for $t < 70$. Subsequently, it abruptly increases as the contour first develops instability at one location, followed by another instability that transforms the contour into an elongated two-fold symmetric shape, similar to the one shown in the middle panel of Fig. 2. Fig. 8c displays the system's energy (2.4) as a function of time. The energy of the circular state is set to zero, yet fluctuations cause a slight elevation in energy. Morphological changes correlate with a sudden decrease in energy succeeded by its increase. The energy surpasses that of the circular state, even though the contour retains an overall elongated shape. This behavior indicates that entropy plays a dominant role in the dynamics. Specifically, for $t > 600$, where the energy saturates, the system stochastically explores the extensive configurational space near the two-fold symmetric state, leading to a slight reduction in the shape parameter.

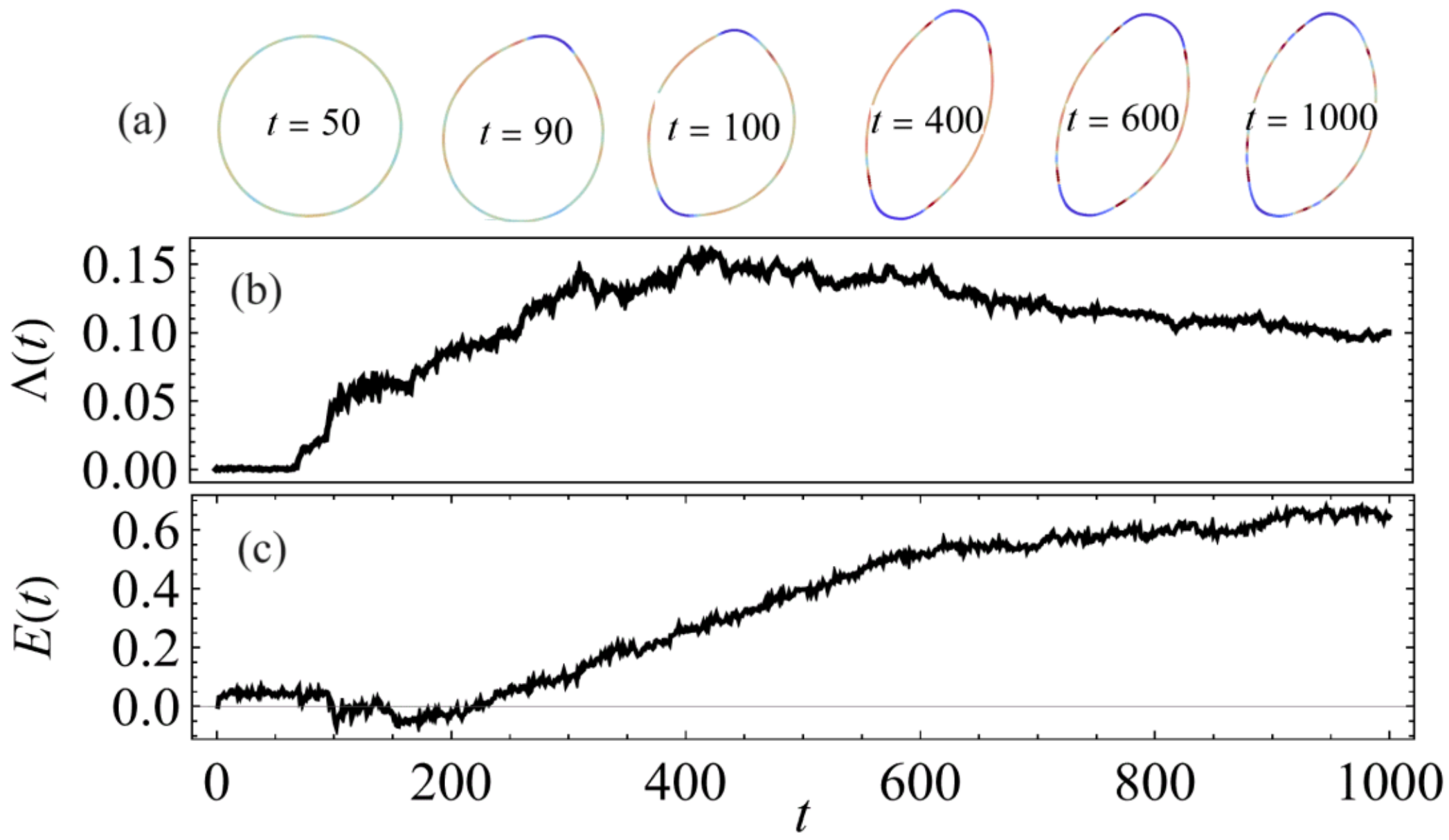

**Figure 8. Time evolution of the contour configurations.** Monte Carlo simulations for the system with parameters: $\kappa = 0.1$, $\eta = 0.5$, short wavelength cutoff $P/30$, and temperature $T = 10^{-3}$. Time units are $\Delta t = N \ln(N)$ Monte Carlo updates, where $N = 300$ is the number of points along the contour used for its discrtization. The initial condition is the unit circle. (a) The contour shape at a few representative time points. The contour colours represent the intensity of the field $\phi$, using a thermometer color code – red and blue for high and low intensity, respectively. (b) Time trace of the shape parameter (9.6). (c) Time trace of the system's energy (2.4). For the time evolution of this system, see Movie 1.

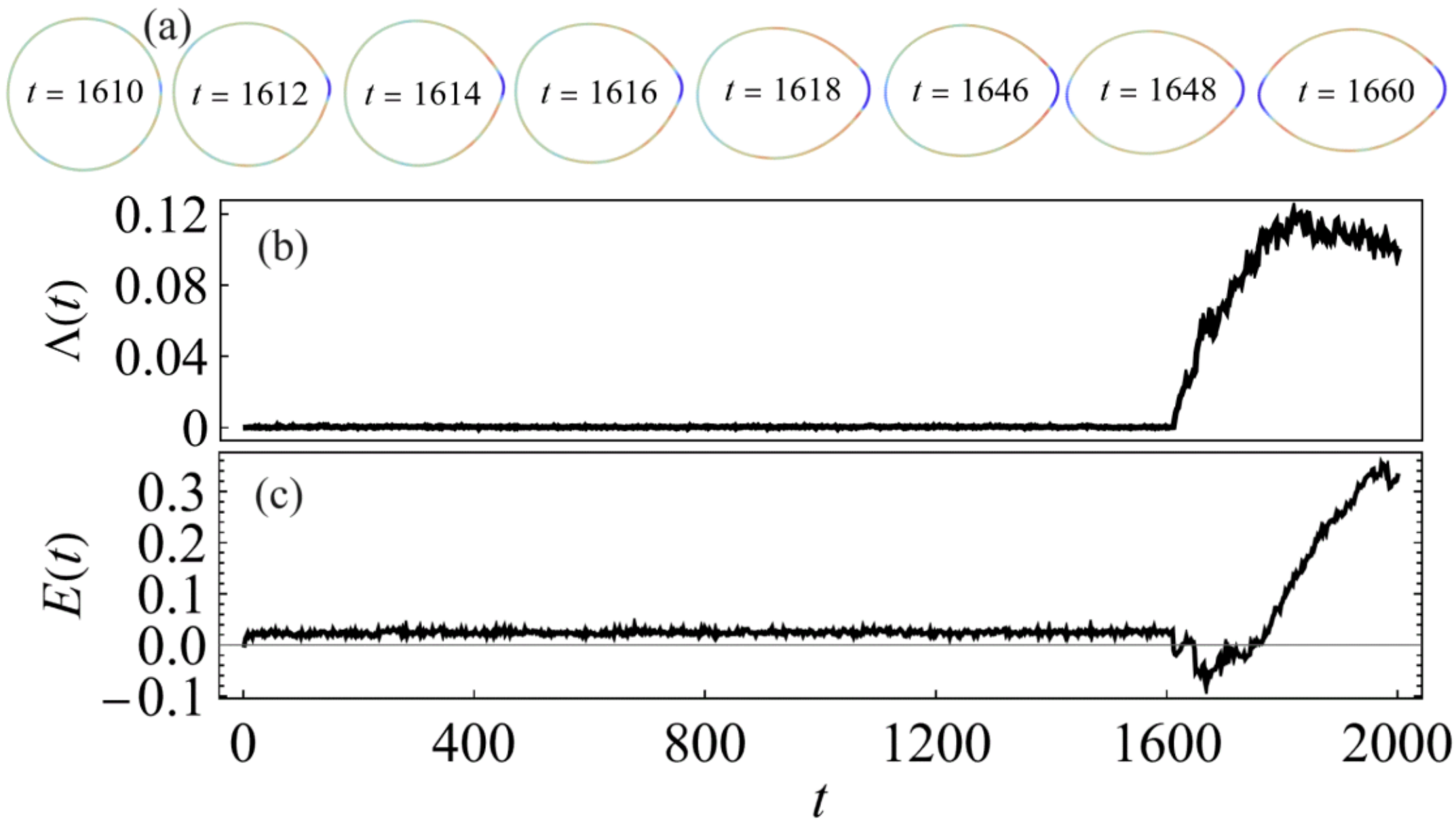

**Figure 9. Time evolution of the contour configurations – Influence of the temperature**. Monte Carlo simulations for the system with parameters: $\kappa = 0.1$, $\eta = 0.5$, short wavelength cutoff $P/30$, and temperature $T = 0.5 \cdot 10^{-3}$. Time units are $\Delta t = N \ln(N)$ Monte Carlo updates, where $N = 300$ is the number of points along the contour used for its discrtization. The initial condition is the unit circle. (a) The contour shape at a few representative time points. (b) Time trace of the shape parameter. (c) Time trace of the system's energy. For time evolution, see Movie 2.

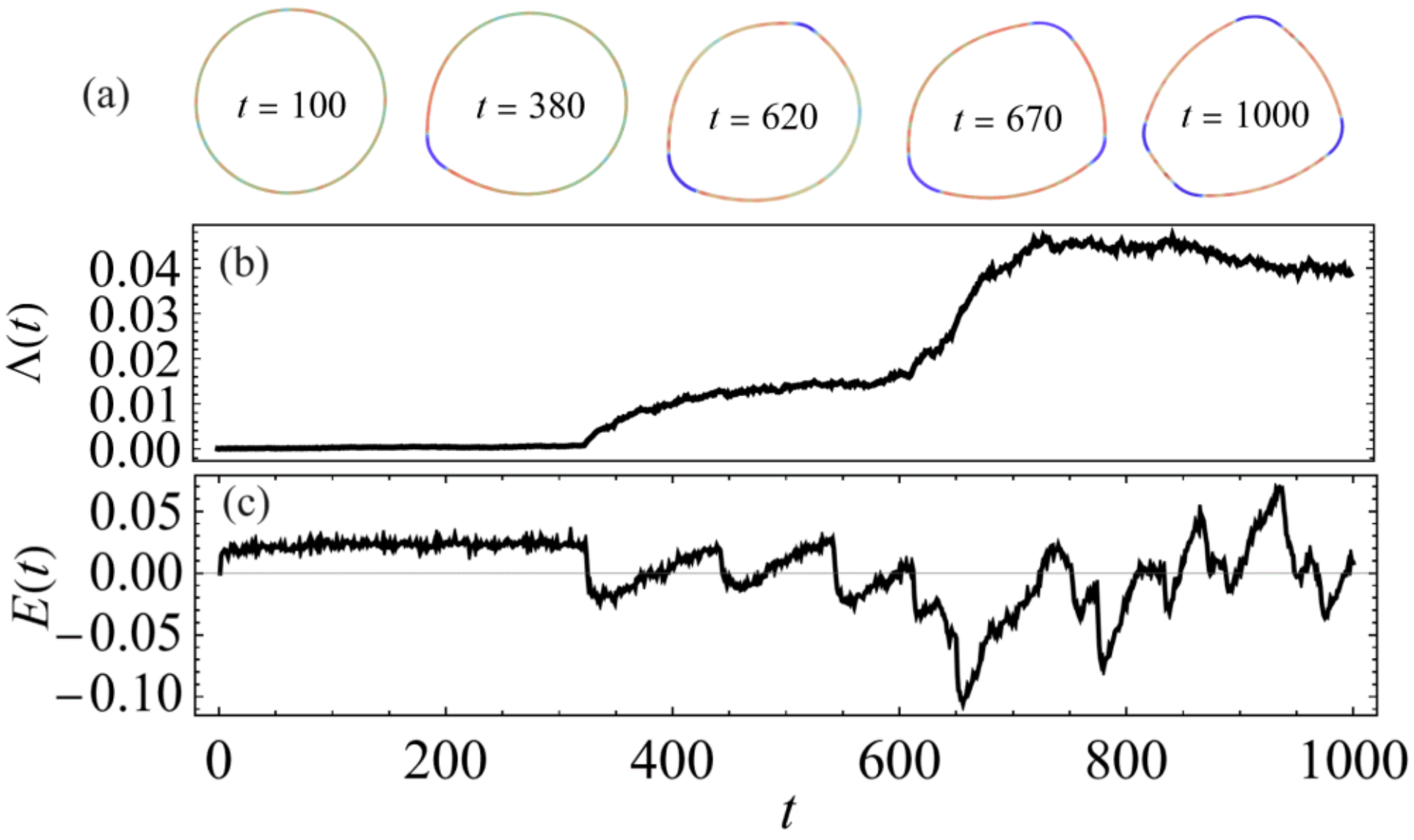

**Figure 10. Time evolution of the contour configurations – The effect of short wavelength cutoff**. Monte Carlo simulations for system parameters: $\kappa = 0.1$, $\eta = 0.5$, short wavelength cutoff $P/60$, and temperature $T = 10^{-3}$. Time units are $\Delta t = N \ln(N)$ Monte Carlo updates, where $N = 300$ is the number of points along the contour used for its discrtization. The initial condition is the unit circle. (a) The contour shape at a few representative time points. (b) Time trace of the shape parameter. (c) Time trace of the system's energy. For time evolution, see Movie 3.

To demonstrate that the transition from the circular state to the elongated one occurs via activation over a barrier, we repeat the calculation of the system's time evolution using the same parameters as in Fig. 8, except that the temperature is reduced by a factor of two, $T = 0.5 \cdot 10^{-3}$. Given that activation exhibits an Arrhenius law behavior, such a temperature reduction is anticipated to result in a significantly extended time period until the transition occurs. Fig. 9 illustrates this characteristic.

Next, we investigate the system's behavior as the lower wavelength cutoff is decreased. A reduction in the lower wavelength cutoff encourages the simultaneous emergence of instability at multiple locations along the contour, thereby increasing the likelihood of the system transitioning into a long-lived metastable state. Fig. 10 presents the results of the Monte Carlo simulations using the same parameters as in Fig. 8, but with a wavelength cutoff reduced by a factor of two, i.e. $P/60$. It illustrates that the system develops instabilities at various points, ultimately transitioning into an approximate four-fold symmetric state. Further reducing the cutoff typically leads to metastable morphologies with more corners.

Finally, we study the case where the coupling parameter surpasses the threshold value (3.9). Here, we consider $\kappa = 0.1$, $\eta = 0.8$, and a temperature that is three orders of magnitude smaller than that of the previous examples, $T = 10^{-6}$. Fig. 11 displays a few examples of the system's behavior within this regime, varying the lower wavelength cutoff. These instances illustrate that, depending on the cutoff value, the system transitions into different metastable states. The energy of these states decreases as the number of corners diminishes, as evidenced by the saturated energy value depicted using the red curve.

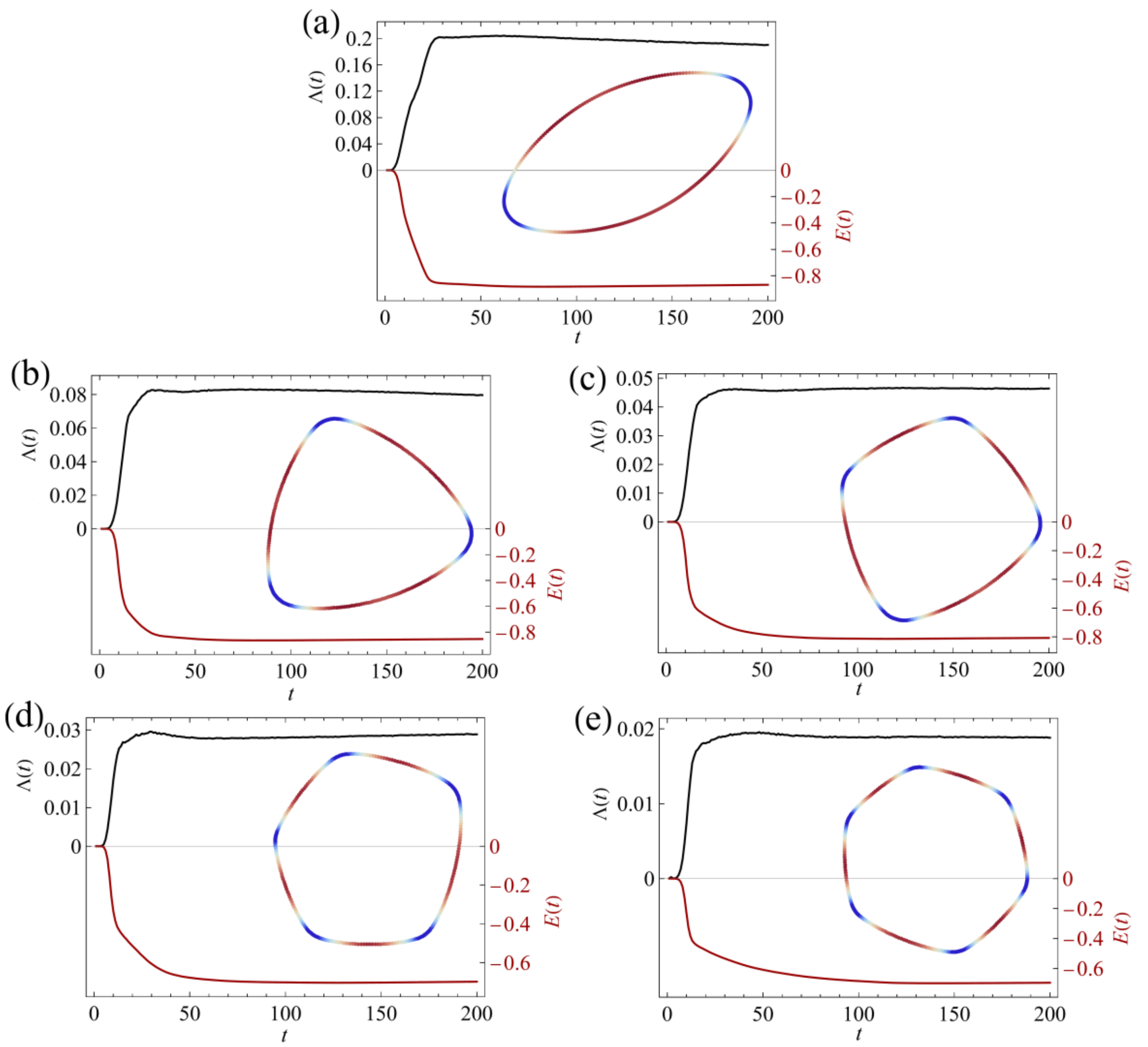

**Figure 11. Time evolution of the contour configuration: Impact of the coupling parameter.** Monte Carlo simulations for system parameters: $\kappa = 0.1$, $\eta = 0.8$, and temperature $T = 10^{-6}$. Time units are $\Delta t = N \ln(N)$ Monte Carlo updates, where $N = 300$ is the number of points along the contour used for its discrtization. The initial condition is the unit circle. The lower wavelength cutoff is $P/b$ where $P$ is the contour perimeter and $b =$ 12, 17.6, 21.4, 20, and 25 for panels (a) (Movie 4a), (b) (Movie 4b), (c) (Movie 4c) ,(d) (Movie 4d), and (e) (Movie 4e), respectively. The black curves show the order parameter while the red ones depict the energy of the system as a function of (the Monte Carlo) time.

## 10. Discussion

We presented a simplified model that captures the essence of a morphological transition of developing biological tissues shifting from a circular to an elongated shape, possibly allowing for various metastable states. Our model outlines various pathways for morphological transitions as the coupling between the scalar field and the contour curvature, $\eta$, increases slowly over time. These pathways are contingent upon the noise intensity (quantified by an effective temperature) and the height of the morphological barrier. The latter separates the minimum of the circular shape of the contour from those minima associated with broken symmetry states. It arises from the presence of an effective lower cut-off on the wavelength of the Fourier components describing the contour shape. Practically, this cutoff could result from the tissue being discretized into individual cells or from mechanisms that limit high curvature values of the tissue's shape.

When the barrier height is negligible relative to the effective temperature, the transition from the circular state to a broken symmetry state is a second-order-like phase transition, while in the opposite limit, the transition involves an activation over a barrier akin to a first-order-like phase transition. In this scenario, stochastic fluctuations are not merely background noise, adding stochasticity to the dynamics; they assume a crucial role in determining the characteristics of the morphogenesis dynamics.

Stochastic fluctuations further introduce entropic barriers within the high-dimensional space of morphological configurations. These barriers arise due to the multitude of trajectories exploring a significant portion of the available phase space of the system. In addition, fluctuations renormalize the energies of the metastable configurations of the system as described in Appendix B. Our Monte Carlo simulations confirm the significant impact of these fluctuations, primarily evident in impeding transitions between metastable states. However, the evolution of the instability from the initial circular configuration into a broken symmetry shape is primarily influenced by the presence of a spatial frequency cutoff on the contour shape. Specifically, the system tends towards the two-fold broken symmetry solution when the frequency cutoff is low, whereas higher-fold broken symmetric states become more probable with an elevated frequency cutoff. The former scenario resembles our experimental observations in *Hydra* regeneration from tissue segments of moderate size [34].

The simple model introduced here represents a novel approach to pattern formation in biological systems. It can serve as a prototype that can be extended to describe more complex systems and the emergence of more intricate morphologies. These extensions may involve the inclusion of multiple fields and diverse compounds within the organism's structure. Drawing parallels with the protein folding problem offers insights into some elements necessary for improving the model. Protein folding is a paradigmatic example of a pattern formation process driven by fluctuations and characterized by numerous entropic barriers in the high-dimensional configuration space. However, assuming an unbiased random search in the configuration space leads to unrealistically long timescales of convergence compared to experimental observations, a dilemma known as Levinthal's paradox [45, 46]. Various mechanisms have been proposed to address this issue, such as forming critical nuclei or folding cores, funnels, hierarchical folding involving secondary structures, and parallel folding pathways. The current understanding of the protein folding problem suggests that a real protein cannot be regarded as a random chain molecule but instead possesses inherent structures that facilitate rapid folding; for a review, see [49]. Similar concepts should be developed and integrated into models describing fluctuation-driven morphological pattern formation, allowing enhanced pattern robustness and acceleration of the dynamics.

Despite its simplicity, our model captures the fundamental aspects of the prominent morphological transition observed in a whole-body *Hydra* regeneration from a small tissue segment - the transition from the incipient spherical shape to a cylindrical one [34]. The key components of the model mirror those found in the *Hydra's* tissue: a scalar calcium field that activates the internal force generation, an internal fluid pressure within the cavity of the enclosed hollow tissue, and the local curvature of the tissue, which influences its morphology. Beyond the major morphological transition, the model offers two main predictions: The first suggests that in *Hydra* regeneration, the order of the morphological transition might change depending on the size of the initial tissue segment. Our previous experiments demonstrated a first-order-like transition in *Hydra* regenerating from moderate-size tissues [34]. The model, however, predicts that the morphological transition should resemble more closely a second-order phase transition in large tissue segments (resulting in an elevated effective upper spatial frequency cut-off) due to the decrease in the height of the morphological barrier. Our recent preliminary experimental results appear to confirm this prediction (work in progress).

The second prediction is that under conditions of strong coupling between the calcium field and tissue's curvature, combined with low levels of fluctuations, the tissue will explore a multitude of metastable morphologies reminiscent of the *n*-fold symmetric configurations in the model. Our experimental setup allows us to manipulate noise intensity by regulating the calcium fluctuations [35]. Exploring experimental approaches to enhance the coupling between curvature and calcium activity could potentially unveil a diverse range of metastable morphologies within the appropriate parameter regime. These findings are expected to stimulate further research into controlling biological pattern formation. They will also deepen our understanding of the underlying physics involved in fluctuation-driven pattern formation across various biological systems, extending beyond the specific case of *Hydra.*

**Acknolegments**

This work was supported by a grant (EB) from the Israel Science Foundation (Grant No. 1638/21)).

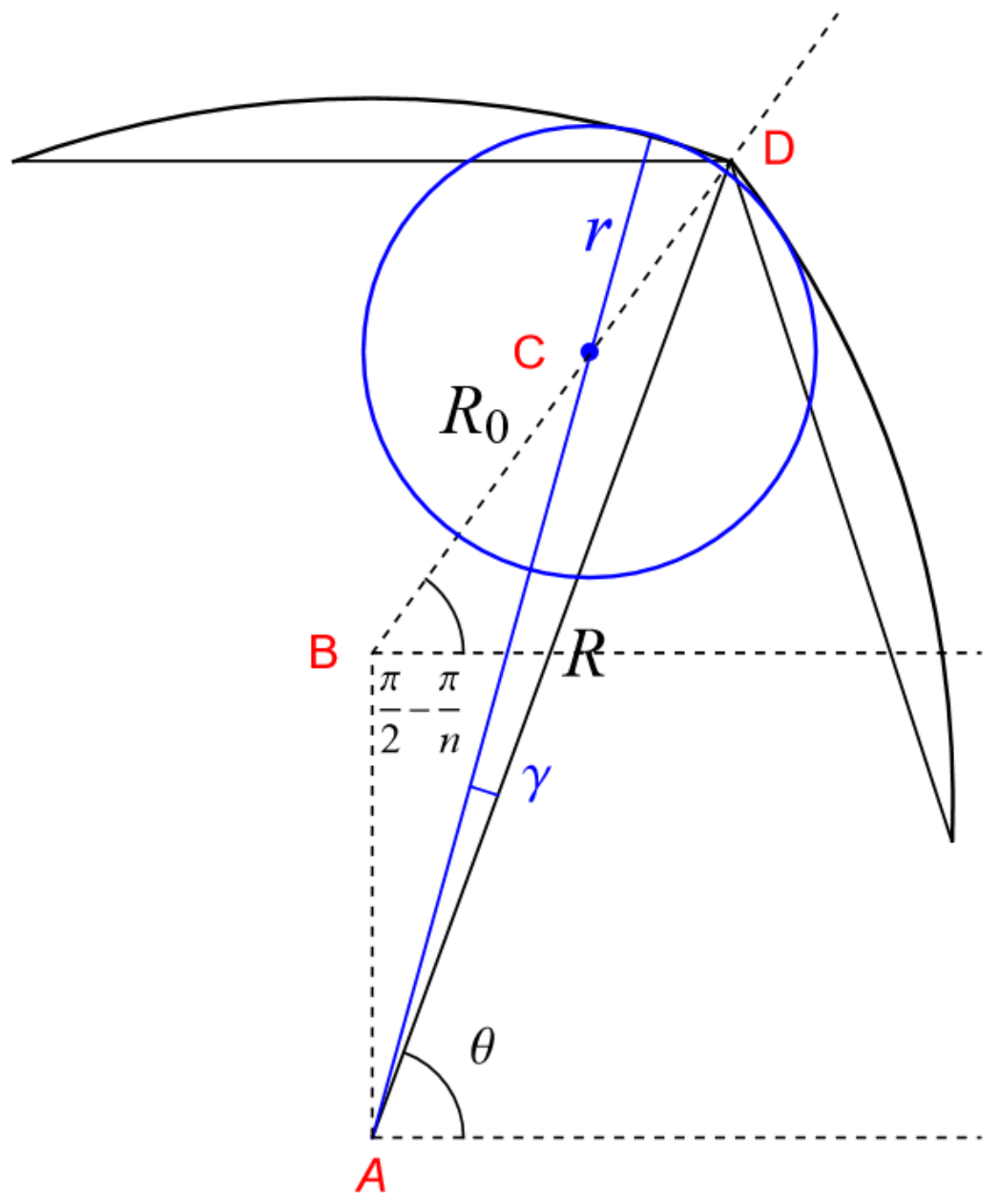

**Figure A1. The geometry of the *n*-fold symmetric configurations of the contour.** Point A corresponds to the center of the circle associated with the larger upper arc, whose radius is $R$ ; Point B represents the center of the *n*-fold symmetric curve; Point C corresponds to the center of the circle associated with the smaller arc with radius $r$ ; Point D designates the intersection point of two larger arcs, with the distance from this point to point B denoted as $R_0$ . For a given value of *n*, the angles $\theta$ and $\gamma$ uniquely define the contour shape subject to the constraint that it encloses an area equal to $\pi$ .

## Appendix A: The *n*-fold symmetric configurations

In this appendix, we extend the computation introduced in section 4 to determine the energy and configurations of contours composed of arcs featuring *n*-fold symmetry, as depicted in Fig. 3. These contours comprise two distinct radii, and their geometric configuration is shown in Fig. A1. Three parameters define them: *n*- the number of long arcs and the two angles, $\theta$ and $\gamma$ .

To compute the energy associated with these contours, we first represent the radii of the two arcs as functions of the angles $\theta$ and $\gamma$ , along with the distance from the origin to the polygon's corner, denoted by $R_0 = \overline{BD}$ . By employing the sine theorem to the triangles $ABD$ and $ACD$, we derive the following relationships:

$$R = \frac{\sin\left(\frac{\pi}{n}\right)}{\sin\left(\frac{\pi}{2}-\theta\right)} R_0 , \quad \text{and} \quad r = \left[1 - \frac{\cos\left(\frac{\pi}{n}+\theta\right)}{\cos\left(\frac{\pi}{n}+\theta+\gamma\right)}\right] R . \tag{A.1}$$

The area enclosed by the contour is given by

$$A = A_{\text{polygon}} + A_{\text{arcs}} - A_{\text{rounding}} , \tag{A.2}$$

where

$$A_{\text{ploygon}} = \frac{n}{2} R_0^2 \sin\left(\frac{2\pi}{n}\right) \tag{A.3}$$

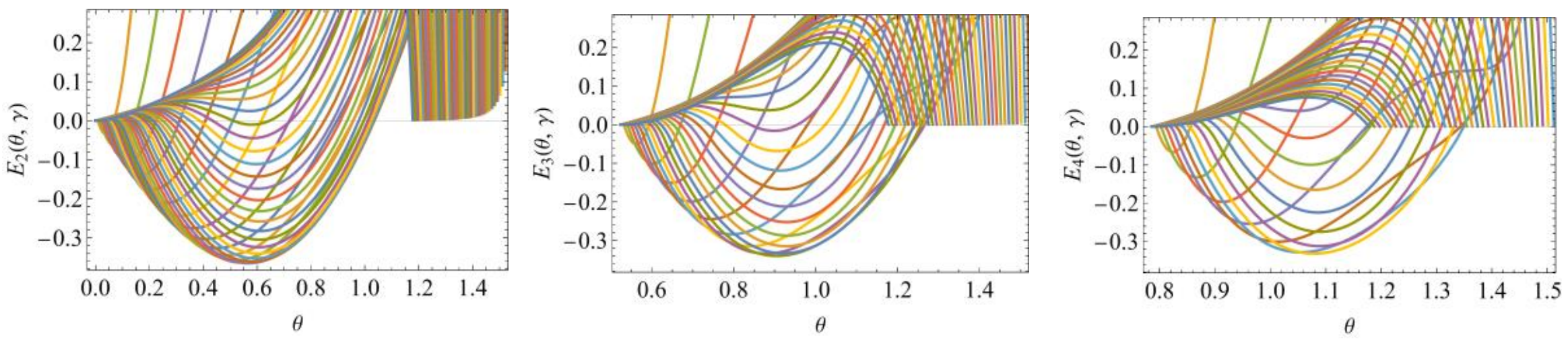

**Figure A2. The energy landscape of high-symmetry configurations.** The plots show the energy $E_n(\theta,\gamma)$ for $n = 2,3$ and 4, as a function of $\theta$ for various values of $\gamma$. In all cases for $\kappa = 0.1$ and $\eta = 0.5$.

is the polygon area, while

$$A_{\text{arcs}} = n\frac{R^2}{2}\left[\pi - 2\theta - \sin(2\theta)\right] \tag{A.4}$$

is the area enclosed between the large arcs and the polygon. Therefore, $A_{\text{polygon}} + A_{\text{arcs}}$ is the enclosed area within the contour in the absence of corner rounding, specifically when $r = 0$. To account for the rounding effect near the corners, it is necessary to subtract the area encompassed by the small osculating circle and the arcs intersecting at the polygon's corner. This area is expressed as:

$$A_{\text{rounding}} = nR^2\left[\gamma - \left(1-\frac{r}{R}\right)\sin\gamma - \frac{r^2}{R^2}\left(\gamma + \pi + \frac{\pi}{n} - \frac{\pi}{2}\right)\right] \tag{A.5}$$

Enforcing the area constraint, $A = \pi$, determines the value of $R_0$.

The total length of the large and the small radii arcs are, respectively,

$$L = nR(\pi - 2\theta - 2\gamma) \quad \text{and} \quad l = 2nr\left(\gamma + \theta + \frac{\pi}{n} - \frac{\pi}{2}\right). \tag{A.6}$$

Contour segments with different curvatures are characterized by two possible energy densities, each dependent on the curvature. Therefore, the system's total energy is the sum of these two energy densities, multiplied by its corresponding arc length, *l* or *L*. For example, the total bending energy of the segments with a large radius of curvature is given by $\kappa(H-1)^2 L = \kappa(1/R-1)^2 L$, while for segments with a small radius of curvature, it is $\kappa(1/r-1)^2 l$. Similarly, the energies associated with the scalar field are $L(\phi_1-1)^2$ and $l(\phi_2-1)^2$, where $\phi_1$ and $\phi_2$ represent the field values on the parts of the contour with large and small radii of curvature, respectively. Taking into account also the field-curvature coupling terms, the the energy for a contour with *n*-fold symmetry is:

$$E_n(\theta,\gamma) = \kappa L\left(\frac{1}{R}-1\right)^2 + \kappa l\left(\frac{1}{r}-1\right)^2 + \frac{\eta L}{R^2}\phi_1 + \frac{\eta l}{r^2}\phi_2 + L(\phi_1-1)^2 + l(\phi_2-1)^2 \tag{A.7}$$

with the optimal values of the fields (that minimize the energy) given by:

$$\phi_1 = \max\left(0, 1-\frac{\eta}{2R^2}\right) \quad \text{and} \quad \phi_2 = \max\left(0, 1-\frac{\eta}{2r^2}\right). \tag{A.8}$$

Examples for $E_n(\theta,\gamma)$, with $n=2,3$ and 4 for $\kappa=0.1$ and $\eta=0.5$ are presented in Fig. A2. They are plotted as a function of $\theta$ and each curve corresponds to a different value of $\gamma$. Notice that the circular state corresponds to the value of $\theta=\pi/2-\pi/n$, hence marking the starting point on the $\theta$ axis.

Tables 1, 2, and 3 provide the results obtained from the minimization of energy $E_n(\theta,\gamma)$, spanning different values of n, ranging from 2 to 10, while maintaining a fixed value of $\kappa=0.1$, and for three coupling parameters: $\eta=$ 0.5, 1, and 1.5. These tables show that under conditions of weak coupling ( $\eta=0.5$ ), the lowest energy state corresponds to a two-fold symmetric configuration. By contrast, when the coupling is strong $\eta=1.5$, configurations with higher symmetries, characterized by $n\geq 3$, attain lower energy levels. It's essential to note that the energy disparity between these configurations is marginal. Consequently, under circumstances of low temperature, the system's dynamics flows to one of these metastable states and it resides there for an extended duration.

**Table 1.** The energy difference between the *n*-fold symmetric state and the circular state, together with the angles that determine the contour shape. The calculation is for $\kappa=0.1$ and . $\eta=0.5$

| $n$ | $-E_n$ | $\theta_{\min}$ | $\gamma_{\min}$ |
|---|---|---|---|
| 2 | 0.365 | 0.562 | 0.135 |
| 3 | 0.340 | 0.904 | 0.088 |
| 4 | 0.333 | 1.072 | 0.065 |
| 5 | 0.330 | 1.172 | 0.052 |
| 6 | 0.329 | 1.239 | 0.043 |
| 7 | 0.328 | 1.287 | 0.037 |
| 8 | 0.327 | 1.327 | 0.033 |
| 9 | 0.327 | 1.350 | 0.028 |
| 10 | 0.327 | 1.372 | 0.026 |

**Table 2.** The same as Table 1 $\kappa=0.1$ and . $\eta=1$

| $n$ | $-E_n$ | $\theta_{\min}$ | $\gamma_{\min}$ |
|---|---|---|---|
| 2 | 1.8151 | 0.929 | 0.158 |
| 3 | 1.8132 | 1.204 | 0.088 |
| 4 | 1.8127 | 1.309 | 0.062 |
| 5 | 1.8125 | 1.365 | 0.048 |
| 6 | 1.8124 | 1.401 | 0.039 |
| 7 | 1.8124 | 1.426 | 0.033 |
| 8 | 1.8123 | 1.445 | 0.029 |
| 9 | 1.8123 | 1.459 | 0.026 |
| 10 | 1.8123 | 1.471 | 0.023 |

**Table 3.** The same as Table 1 $\kappa=0.1$ and . $\eta=1.5$

| $n$ | $-E_n$ | $\theta_{\min}$ | $\gamma_{\min}$ |
|---|---|---|---|
| 2 | 2.7859 | 1.064 | 0.155 |
| 3 | 2.8224 | 1.313 | 0.073 |
| 4 | 2.8303 | 1.391 | 0.049 |
| 5 | 2.8334 | 1.432 | 0.037 |
| 6 | 2.8350 | 1.457 | 0.030 |
| 7 | 2.8359 | 1.474 | 0.025 |
| 8 | 2.8365 | 1.486 | 0.022 |
| 9 | 2.8369 | 1.496 | 0.019 |
| 10 | 2.8372 | 1.504 | 0.017 |

## Appendix B: Linear stability analysis of the *n*-fold symmetric configurations

This appendix is dedicated to examining the stability of *n*-fold symmetric patterns of the contour. We leverage our findings to conduct a renormalization of their energies. To embark on this task, our initial step is to identify the polar representation of the *n*-fold symmetric curve, denoted as $\rho_n^{(0)}(\varphi)$, with $0 \le \varphi < 2\pi$. Throughout our analysis, we assume that the orientation of the *n*-fold symmetric curve aligns such that one of its long edges is parallel to the $x$-axis, exemplified by the instances depicted in Fig. 3. Then the equation for the upper long arc sector is

$$x^2 + (y+h)^2 = R^2 \tag{B.1}$$

with

$$h = R\sin\theta - R_0 \sin\left(\frac{\pi}{2} - \frac{\pi}{n}\right) = R\sin\theta - R_0 \cos\left(\frac{\pi}{n}\right) \tag{B.2}$$

where $R$ and $R_0$ are defined in Fig. A1. Substituting

$$x = \rho_n^{(0)}(\varphi)\cos\varphi \text{ and } y = \rho_n^{(0)}(\varphi)\sin\varphi \tag{B.3}$$

in (B.1) and solving for $\rho_n^{(0)}(\varphi)$ we obtain the polar representation of the long arc:

$$\rho_n^{(0)}(\varphi) = -h\sin\varphi + \sqrt{R^2 - h^2\cos^2\varphi}\ . \tag{B.4}$$

The validity range of this expression is $\varphi_* < \varphi < \pi - \varphi_*$, where the angle $\varphi_*$ can be deduced from Fig. A1 to be

$$\varphi_* = \tan^{-1}\left(\frac{R\sin(\theta+\gamma) - h}{R\cos(\theta+\gamma)}\right). \tag{B.5}$$

For the short arc, the contour equation takes the form

$$(x-x_0)^2 + (y-y_0)^2 = r^2, \tag{B.6}$$

where the coordinates of the circle's center are

$$\begin{aligned} x_0 &= (R-r)\cos(\theta+\gamma), \\ y_0 &= (R-r)\sin(\theta+\gamma) - h. \end{aligned} \tag{B.7}$$

Substituting (B.3) in (B.6) and solving for $\rho_n^{(0)}(\varphi)$ we obtain:

$$\begin{aligned} \rho_n^{(0)}(\varphi) &= (R-r)\cos(\theta+\gamma-\varphi) - h\sin\varphi \\ &+ \sqrt{\left[(R-r)\cos(\theta+\gamma-\varphi) - h\sin\varphi\right]^2 - R^2 + 2Rr - h^2 + 2h(R-r)\sin(\theta+\gamma)} \end{aligned} \tag{B.8}$$

which is valid for

$$0 < \varphi < \varphi_*. \tag{B.9}$$

Finally, in the regime $\pi - \varphi_* < \varphi < \pi$ the polar representation is given by $\rho_n^{(0)}(\pi - \varphi)$.

To look for the stability characteristics of the configuration, we now add a perturbation $\delta\rho_n(\varphi)$ to the curve expressed as a Fourier series with arbitrary expansion coefficients, i.e. $\rho_n(\varphi)=\rho_n^{(0)}(\varphi)+\delta\rho_n(\varphi)$ with

$$\delta\rho_n(\varphi)=c_0+\sum_{j=2}^{\infty}c_j\cos\left[j(\varphi-\beta_n)\right]+d_j\sin\left[j(\varphi-\beta_n)\right], \tag{B.10}$$

where

$$\beta_n=\begin{cases}0 & \text{even } n\\ \dfrac{\pi}{2n} & \text{odd } n\end{cases} \tag{B.11}$$

The reason for adding this phase shift will become apparent below. Notice that this Fourier expansion excludes the $j=1$ which is associated with a trivial rigid translation of the contour.

The area constraint

$$\frac{1}{2\pi}\int_0^{2\pi}d\varphi\left[\rho_n(\varphi)\right]^2=1, \tag{B.12}$$

is imposed through the division of the perturbed polar representation by the proper normalization constant:

$$\rho_n(\varphi)\to\bar{\rho}_n(\varphi)=\frac{\rho_n(\varphi)}{\sqrt{1+2c_0a_0+c_0^2+\sum_{j=2}^{\infty}a_jc_j+\frac{1}{2}\sum_{j=2}^{\infty}\left(c_j^2+d_j^2\right)}}, \tag{B.13}$$

where $a_j$ and $b_j$ are the Fourier expansion coefficients of the unperturbed contour,

$$\rho_n^{(0)}(\varphi)=a_0+\sum_{j=1}^{\infty}a_{nj}\cos\left[nj(\varphi-\beta_n)\right]. \tag{B.14}$$

The choice of the above phase shift allows us to expand the polar representation of the unperturbed curve using only cosine functions.

The energy functional $E\left[\bar{\rho}_n(\varphi)\right]$, after minimization over the scalar field $\phi$, is a function of the sets of coefficients $\{c_j\}$ and $\{d_j\}$. The Hessian matrix is constructed from the second derivatives with respect to these coefficients:

$$\mathcal{H}_{ij}=\left.\frac{\partial^2E\left[\bar{\rho}_n(\varphi)\right]}{\partial g_i\partial g_j}\right|_{\{g_i\}=0}, \tag{B.15}$$

where $g_i$ is one of the components of the vector whose entries are the Fourier coefficients:

$$\boldsymbol{g}^T=\left(c_0,c_2,c_3\cdots,d_2,d_3,\cdots\right). \tag{B.16}$$

The eigenvalues of $\mathcal{H}$ characterize the stability of the corresponding contour shape, and the eigenmode of the least stable is the mode that features the strongest fluctuations when the temperature is finite.

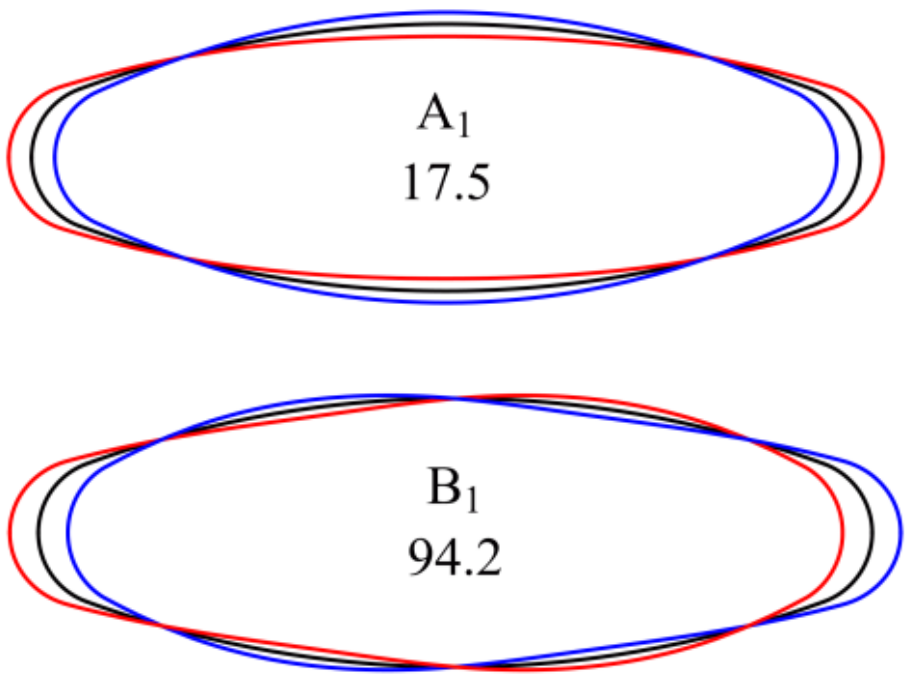

**Figure B1.** Small deformation modes of the two-fold symmetric configuration classified by irreducible representations. The presented modes correspond to the lowest eigenvalue of the Hessian matrix associated with each irreducible representation.

Considerable reduction in the computational effort required for diagonalizing the Hessian matrix can be achieved by exploiting the contour symmetry. Specifically, the Hessian matrix can be transformed into a block diagonal structure, where each block corresponds to an irreducible representation of the associated symmetry group. In other words, the perturbation (B.10) can be expanded in the basis functions of one of the irreducible representations belonging to the point group associated with the *n*-fold symmetric state. Subsequently, the resulting sector of the Hessian matrix is subjected to diagonalization. For instance, when dealing with a 2-fold symmetric contour, the corresponding point group $C_{2V}$ possesses four one-dimensional irreducible representations: A1, A2, B1, and B2. These representations are associated, respectively, with basis functions denoted as follows: $\{\cos(2k\varphi)\}_{k=0}^{\infty}$, $\{\sin(2k\varphi)\}_{k=1}^{\infty}$, $\{\cos((2k+1)\varphi)\}_{k=1}^{\infty}$, and $\{\sin((2k+1)\varphi)\}_{k=1}^{\infty}$. The modes associated with the lowest eigenvalue of each irreducible representation are presented in Fig. B1 for $\kappa = 0.1$ and $\eta = 1.5$. These modes are derived through the diagonalization of the Hessian matrix, utilizing a truncated basis comprising the 100 lowest basis functions associated with the given irreducible representation. Based on the results obtained for a smaller set of basis functions, we estimate the eigenvalues' error to be less than 5%.

Similarly, the point group $C_{3v}$, associated with the 3-fold symmetric curve, has two one-dimensional irreducible representations, $A_1$ and $A_2$, as well as a single two-dimensional representation E, with basis functions respectively given by $\{\cos(3k\phi)\}_{k=0}^{\infty}$, $\{\sin(3k\phi)\}_{k=1}^{\infty}$, and all other harmonics. Fig. B2 shows the deformation modes of the lowest eigenvalues of the Hessian matrix and the fundamental deformation modes of 4-fold, 5-fold, and 6-fold symmetric curves, all computed using the same parameters: $\kappa = 0.1$ and $\eta = 1.5$.

Notably, the least stable deformation modes are consistently associated with the elongation of the *n*-fold symmetric state. This tendency is particularly conspicuous in the B2 mode for 4-fold symmetric, the E2 mode for 5-fold symmetric, and the E2 mode for 6-fold symmetric configurations.

The procedure for calculating eigenvalues of the Hessian and their corresponding modes involves the diagonalization of a matrix derived from the expansion (B.10), incorporating a cutoff $j_{\max} \geq 200$. Based on the analysis of smaller matrices, the error estimation is less than 5%. A log-log plot displaying the lowest 120 eigenvalues, which we denote by $\lambda_k^{(n)}$, is depicted in Fig. B3. The asymptotic power-law behavior at large wave numbers, $\lambda_k^{(n)} \sim k^4$, can be understood using the following rationale. The energy associated with short-wavelength deformations is essentially unaffected by the specific contour configuration, rendering its wavelength dependence analogous to that of a circular contour where it can be readily deduced from Equation (3.8) that the energy cost scales as $k^4$.

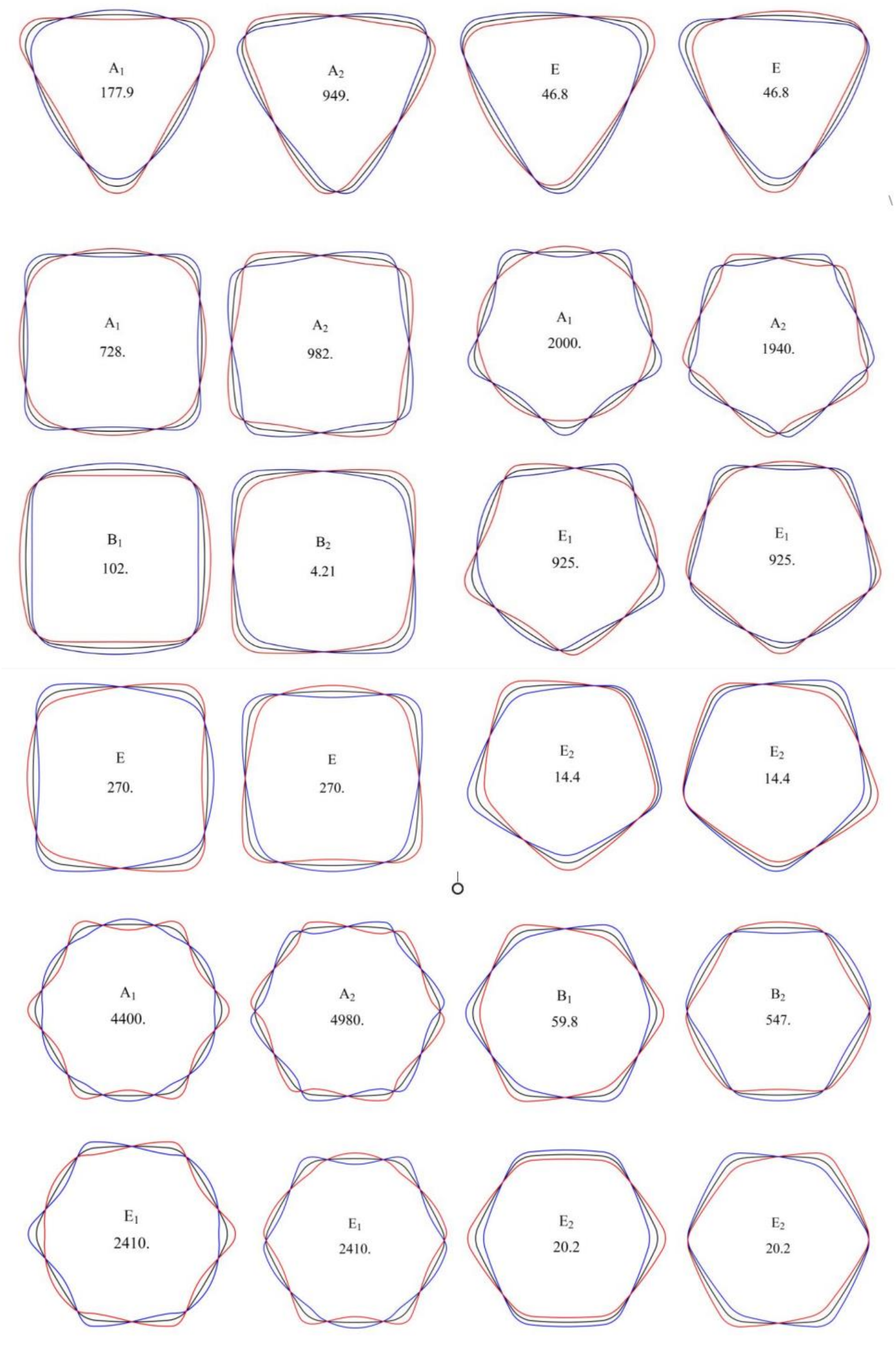

**Figure B2.** Deformation modes of the *n*-fold symmetric configurations, with $3 \le n \le 6$ classified by irreducible representations of the corresponding point group.

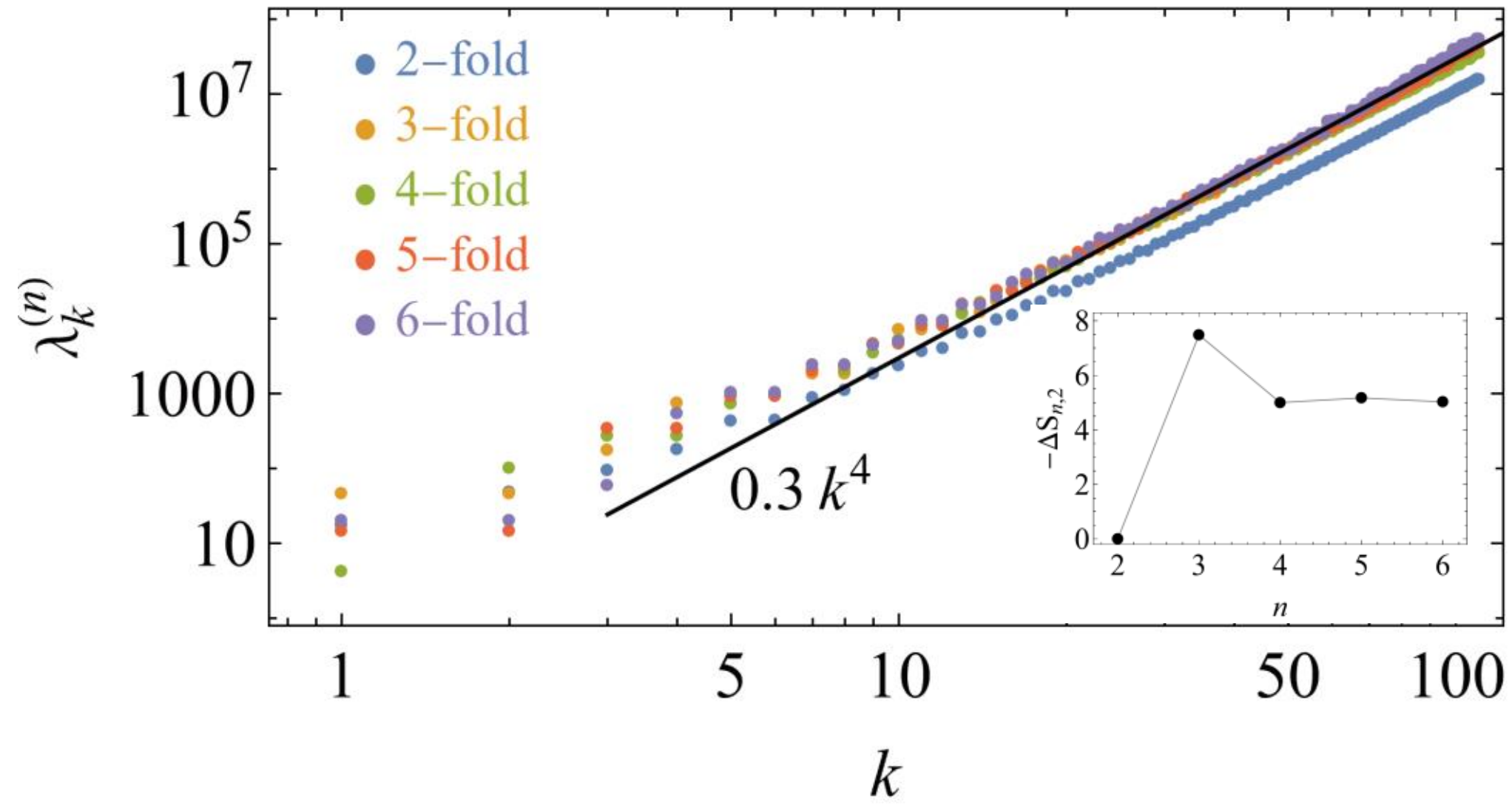

**Figure B3.** A log-log plot of the lowest 120 eigenvalues of the Hessian matrix (B.15) corresponding to the *n*-fold symmetric configurations with $2 \le n \le 6$. Inset: the entropy difference, given by Eq. (B.20), calculated for $k_c = 10$.

The set of eigenvalues of the Hessian matrix associated with the *n*-fold symmetric configuration allows one to calculate the renormalized value of the corresponding energy that takes into account the fluctuations. This renormalized energy is essentially the free energy and is given by:

$$F_n = -T \ln \left\{ \int \prod_{j=1}^{j_c} dg_j \exp\left[ -\frac{1}{T} \left( E_n + \frac{1}{2} \sum_{ij} g_i \mathcal{H}_{ij} g_j \right) \right] \right\}. \tag{B.17}$$

Here, we introduced an upper cutoff on the wavelength of the possible fluctuations of the contour, which we denote by $k_c$. The above Gaussian integral yields

$$F_n = E_n - TS_n, \quad \text{with} \quad S_n = \ln\left( \prod_{k=1}^{k_c} \frac{\sqrt{2\pi T}}{\lambda_k^{(n)}} \right). \tag{B.18}$$

Specifically, the free energy difference between an *n*-fold and the 2-fold symmetric configurations is

$$\Delta F = F_n - F_2 = \Delta E_{n,2} - T\Delta S_{n,2}. \tag{B.19}$$

with

$$\Delta E_{n,2} = E_{n,\min} - E_{2,\min}, \text{ and } \quad \Delta S_{n,2} = \sum_{k=1}^{k_c} \ln\left( \frac{\lambda_k^{(2)}}{\lambda_k^{(n)}} \right). \tag{B.20}$$

The entropy difference, $-\Delta S_{n,2}$, as a function of $n$, computed using the Hessian eigenvalues with a cutoff at $k_c = 10$, is presented in the inset of Fig. B3. It reveals that the highest entropy corresponds to the 2-fold symmetric configuration. Consequently, under conditions of sufficiently elevated temperature, the free energy $F_n$ reaches its minimum value when $n = 2$. In simpler terms, when there are significant fluctuations in the system the most likely contour configuration is an elongated shape (specifically, 2-fold symmetric) with some small temporal deformations.

## Appendix C: The condition for global stability of the circular state

In this appendix, we establish the validity of Equation (6.1) which defines the critical coupling strength $\eta_c$ setting the condition for the global stability of the circular configuration of the contour. Global stability is achieved when the minimum energy of the two-fold symmetric shape is higher than the minimum of the circular state, where $E_{\text{circ.}} = 0$. To find the value of $\eta$ that ensure this, it's necessary to calculate the minimal energy of the two-fold symmetric state. However, as shown in Figure 5, once this energy dips below zero, the energy $E_2(\theta,\gamma)$ minimized over $\theta$, for any fixed value of $\gamma$, also becomes negative, particularly as $\gamma \to 0$. Thus, to identify the critical value of $\eta$, it is sufficient to determine the value for which:

$$\lim_{\gamma \to 0} \min_{\theta} E_2(\theta,\gamma) = 0 \tag{C.1}$$

where the minimization is over the angle $\theta$ while keeping $\gamma$ fixed.

The condition $\gamma \to 0$ enables us to expand the energy $E_2(\theta,\gamma)$ to first order in $\gamma$ and to second order in $\theta$. Specifically, this expansion yields the length and the radius of curvature of the large circular arc as:

$$L = \pi - 2\gamma - \frac{4\gamma\theta}{\pi} + \frac{2(\pi - 6\gamma)\theta^2}{\pi^2} \tag{C.2}$$

and

$$r_L = 1 + \frac{2\theta}{\pi} + \frac{6\theta^2}{\pi^2}, \tag{C.3}$$

while the length and radius of curvature of the small circular arc are:

$$l = \gamma\left(2 + \frac{4\theta}{\pi} + \frac{2(18 - \pi^2)}{3\pi^2}\theta^2\right) \tag{C.4}$$

and

$$r_l = r_L - \frac{\theta(\pi + 2\theta)}{\pi(\gamma + \theta)} - \frac{\theta(\gamma + \theta)(\pi + 8\theta)}{6\pi}. \tag{C.5}$$

To obtain the last expression, we first expanded the curvature in $\gamma$ around $-\theta$ and then expand to second order in $\theta$.

With these expressions and taking into account that the minimal energy occurs $\phi = 1 - \eta/(2r_L^2)$ on the large arc and $\phi = 0$ on the small arc, we obtain:

$$E_2(\theta,\gamma) \simeq 2L\kappa\left(\frac{1}{r_L} - 1\right)^2 + 2l\kappa\left(\frac{1}{r_l} - 1\right)^2 + \frac{\eta}{r_L^2}\left(1 - \frac{\eta}{4r_L^2}\right)2L. \tag{C.6}$$

Calculating the minimum of this energy as a function of $\theta$, and expanding to linear order in $\gamma$, we arrive at:

$$E_{\min} \simeq \frac{\gamma}{\kappa}(\eta-2)^2\left(\eta^2-\kappa\right). \tag{C.7}$$

Hence, the physical solution to Equation (C.1) is given by Equation (6.1).

It is crucial to note that this calculation is valid only when the circular state is locally stable, i.e., when $\eta < \eta_{\text{th}}$. Therefore, Eq. (6.1) holds as long as $\eta_c \le \eta_{\text{th}}$. Substituting the formulas (3.9) and (6.1) into this inequality leads to the condition $\kappa < 4/9$.

## Appendix D: Derivation of the force formulas (8.4) and (8.5)

In this appendix, we derive expressions (8.4) and (8.5) for the forces that appear in the Langevin equation (8.1). The force associated with the $\phi$ field (8.3) can be readily obtained by considering the variation of the energy functional (2.4) with respect to $\phi$. The variation with respect to the contour shape is more problematic and will be described below.

When considering the variation of the energy with respect to the contour deformation, it is essential to acknowledge that such a deformation alters the arclength. Consequently, it becomes more convenient to perform the variation using a fixed parameterization, such as $\boldsymbol{r}(\theta)$, where $\theta$ is the angle between the vector $\boldsymbol{r}$ and an arbitrary fixed axis. Then, after the variation, to transform back to the arclength parametrization.

Thus, we consider variations of the form

$$\boldsymbol{r}(\theta) \to \boldsymbol{r}(\theta) + \delta\boldsymbol{r}(\theta), \tag{D.1}$$

which implies that

$$\boldsymbol{r}_\theta = \frac{d\boldsymbol{r}}{d\theta} \to \boldsymbol{r}_\theta + \frac{d}{d\theta}\delta\boldsymbol{r}, \quad \text{and} \quad \boldsymbol{r}_{\theta\theta} = \frac{d^2\boldsymbol{r}}{d\theta^2} \to \boldsymbol{r}_{\theta\theta} + \frac{d^2}{d\theta^2}\delta\boldsymbol{r}. \tag{D.2}$$

The energy is an integral of the form

$$E = \int d\theta K\left(\boldsymbol{r}_\theta, \boldsymbol{r}_{\theta\theta}\right), \tag{D.3}$$

where the kernel, $K\left(\boldsymbol{r}_\theta, \boldsymbol{r}_{\theta\theta}\right)$, is a function of the "velocity" $\boldsymbol{r}_\theta$ and the "acceleration" $\boldsymbol{r}_{\theta\theta}$, as well as the scalar field which we do not write explicitly. Variation of the energy and integration by parts yields:

$$\delta E = \int d\theta \frac{K\left(\boldsymbol{r}_\theta, \boldsymbol{r}_{\theta\theta}\right)}{\partial \boldsymbol{r}_\theta}\frac{d}{d\theta}\delta\boldsymbol{r} + \frac{\partial K\left(\boldsymbol{r}_\theta, \boldsymbol{r}_{\theta\theta}\right)}{\partial \boldsymbol{r}_{\theta\theta}}\frac{d^2}{d\theta^2}\delta\boldsymbol{r} = \int d\theta\left[\frac{d^2}{d\theta^2}\left(\frac{\partial K\left(\boldsymbol{r}_\theta, \boldsymbol{r}_{\theta\theta}\right)}{\partial \boldsymbol{r}_{\theta\theta}}\right) - \frac{d}{d\theta}\frac{K\left(\boldsymbol{r}_\theta, \boldsymbol{r}_{\theta\theta}\right)}{\partial \boldsymbol{r}_\theta}\right]\delta\boldsymbol{r} \tag{D.4}$$

Hence, the force is given by

$$\tilde{\boldsymbol{f}} = -\frac{\delta E}{\delta \boldsymbol{r}} = -\frac{d^2}{d\theta^2}\left[\frac{\partial K\left(\boldsymbol{r}_\theta, \boldsymbol{r}_{\theta\theta}\right)}{\partial \boldsymbol{r}_{\theta\theta}}\right] + \frac{d}{d\theta}\left[\frac{K\left(\boldsymbol{r}_\theta, \boldsymbol{r}_{\theta\theta}\right)}{\partial \boldsymbol{r}_\theta}\right]. \tag{D.5}$$

The tilde symbol signifies that $\tilde{\boldsymbol{f}}$ is a force per unit angle. It will be later transformed into a force per unit arclength.

In what follows, we calculate this force for the various terms that appear in the energy functional (2.4). Starting with the field term,

$$K = \left|\frac{d\boldsymbol{r}}{d\theta}\right|\left(\phi(\theta)-1\right)^2, \tag{D.6}$$

we obtain:

$$\tilde{\boldsymbol{f}}_\phi = \frac{d}{d\theta}\left[\frac{\partial}{\boldsymbol{r}_\theta}\sqrt{\boldsymbol{r}_\theta\cdot\boldsymbol{r}_\theta}\left(\phi(\theta)-1\right)^2\right] = \frac{d}{d\theta}\left[\frac{\boldsymbol{r}_\theta}{|\boldsymbol{r}_\theta|}\left(\phi(\theta)-1\right)^2\right]. \tag{D.7}$$

To express this force in terms of the arclength parametrization, we use the relation $\tilde{\boldsymbol{f}}_\phi d\theta = \boldsymbol{f}_\phi ds$ which implies that,

$$\boldsymbol{f}_\phi = \frac{d}{ds}\left[\boldsymbol{r}_s\left(\phi(s)-1\right)^2\right] \ . \tag{D.8}$$

Here we use the fact that $\boldsymbol{r}_\theta/|\boldsymbol{r}_\theta|$ is a unit vector tangential to the contour hence it is $\boldsymbol{r}_s$.

Henceforth, we shall represent the tangent vector as $\boldsymbol{t}$, whereas $\boldsymbol{n}$ will denote a unit vector perpendicular to the contour and pointing outwards. These two vectors satisfy the following relations:

$$\frac{d\boldsymbol{t}}{ds} = \boldsymbol{r}_{ss} = -\boldsymbol{n}H \tag{D.9}$$

where $H$ is the curvature, and

$$\frac{d\boldsymbol{n}}{ds} = -\boldsymbol{t}H \ . \tag{D.10}$$

Using (D.9) for calculating the derivative in (D.8) we arrive at:

$$\boldsymbol{f}_\phi = -\boldsymbol{n}H\left(\phi-1\right)^2 + \boldsymbol{t}\frac{d}{ds}\left(\phi-1\right)^2 . \tag{D.11}$$

Consider a kernel which is quadratic in the curvature and linear in the field,

$$K = \left|\frac{d\boldsymbol{r}}{d\theta}\right|H^2\phi \ . \tag{D.12}$$

To express the curvature in terms of the parameter $\theta$ it is convenient to regard the vectors $\boldsymbol{r}_\theta$ and $\boldsymbol{r}_{\theta\theta}$ as three dimensional vectors residing in the $xy$ plane and set $\boldsymbol{z}$ to be a unit vector perpendicular to that plane, such that $\boldsymbol{n}$, $\boldsymbol{t}$ and $\boldsymbol{z}$ form a right-hand coordinate system. With this choice, the curvature is given by

$$H(\theta) = \frac{\boldsymbol{z}\cdot\left(\boldsymbol{r}_\theta\times\boldsymbol{r}_{\theta\theta}\right)}{\left(\boldsymbol{r}_\theta\cdot\boldsymbol{r}_\theta\right)^{3/2}} . \tag{D.13}$$

Thus, the force associated with the kernel given by (D.12) is

$$\tilde{\boldsymbol{f}}_\eta = \frac{d}{d\theta}\frac{\partial}{\partial\boldsymbol{r}_\theta}\left\{\sqrt{\boldsymbol{r}_\theta\cdot\boldsymbol{r}_\theta}\left[\frac{\boldsymbol{z}\cdot\left(\boldsymbol{r}_\theta\times\boldsymbol{r}_{\theta\theta}\right)}{\left(\boldsymbol{r}_\theta\cdot\boldsymbol{r}_\theta\right)^{3/2}}\right]^2\phi\right\} - \frac{d^2}{d\theta^2}\left\{\frac{\partial}{\partial\boldsymbol{r}_{\theta\theta}}\sqrt{\boldsymbol{r}_\theta\cdot\boldsymbol{r}_\theta}\left[\frac{\boldsymbol{z}\cdot\left(\boldsymbol{r}_\theta\times\boldsymbol{r}_{\theta\theta}\right)}{\left(\boldsymbol{r}_\theta\cdot\boldsymbol{r}_\theta\right)^{3/2}}\right]^2\phi\right\} . \tag{D.14}$$

Using the relations

$$\frac{\partial}{\partial \mathbf{r}_\theta} \mathbf{z} \cdot (\mathbf{r}_\theta \times \mathbf{r}_{\theta\theta}) = \frac{\partial}{\partial \mathbf{r}_\theta} \mathbf{r}_\theta \cdot (\mathbf{r}_{\theta\theta} \times \mathbf{z}) = -\mathbf{z} \times \mathbf{r}_{\theta\theta} \tag{D.15}$$

and

$$\frac{\partial}{\partial \mathbf{r}_{\theta\theta}} \mathbf{z} \cdot (\mathbf{r}_\theta \times \mathbf{r}_{\theta\theta}) = \frac{\partial}{\partial \mathbf{r}_{\theta\theta}} \mathbf{r}_{\theta\theta} \cdot (\mathbf{z} \times \mathbf{r}_\theta) = \mathbf{z} \times \mathbf{r}_\theta \tag{D.16}$$

we obtain

$$\tilde{\boldsymbol{f}}_\eta = -\frac{d}{d\theta} \left\{ \frac{2\mathbf{z} \cdot (\mathbf{r}_\theta \times \mathbf{r}_{\theta\theta})}{(\mathbf{r}_\theta \cdot \mathbf{r}_\theta)^{3/2}} \frac{(\mathbf{z} \times \mathbf{r}_{\theta\theta})}{(\mathbf{r}_\theta \cdot \mathbf{r}_\theta)} \phi + 5 \left( \frac{\mathbf{z} \cdot (\mathbf{r}_\theta \times \mathbf{r}_{\theta\theta})}{(\mathbf{r}_\theta \cdot \mathbf{r}_\theta)^{3/2}} \right)^2 \frac{\mathbf{r}_\theta}{|\mathbf{r}_\theta|} \phi + \frac{d}{d\theta} \left[ \frac{2\mathbf{z} \cdot (\mathbf{r}_\theta \times \mathbf{r}_{\theta\theta})}{(\mathbf{r}_\theta \cdot \mathbf{r}_\theta)^{3/2}} \frac{(\mathbf{z} \times \mathbf{r}_\theta)}{(\mathbf{r}_\theta \cdot \mathbf{r}_\theta)} \right] \right\}. \tag{D.17}$$

Next, to express $\boldsymbol{r}_\theta$ and $\boldsymbol{r}_{\theta\theta}$ in terms of the arclength parameter, we use the relation, $ds = |\boldsymbol{r}| d\theta$ which implies that:

$$\begin{aligned} \boldsymbol{r}_\theta &= \frac{\partial s}{\partial \theta} \frac{\partial}{\partial s} \boldsymbol{r} = \rho \boldsymbol{r}_s \\ \boldsymbol{r}_{\theta\theta} &= \rho \frac{\partial}{\partial s} (\rho \boldsymbol{r}_s) = \rho \frac{\partial \rho}{\partial s} \boldsymbol{r}_s + \rho^2 \boldsymbol{r}_{ss} \end{aligned} \qquad \text{with } \rho = \frac{ds}{d\theta}. \tag{D.18}$$

Substituting these relations in (D.17), one can check that the terms that depend on $\rho$ cancel each other (as they should), and we are left with:

$$\boldsymbol{f}_\eta = -\frac{d}{ds} \left\{ 4H(s)(\boldsymbol{z} \times \boldsymbol{r}_{ss})\phi + 5H^2(s)\boldsymbol{r}_s \phi + 2\frac{\partial H\phi}{\partial s} \boldsymbol{z} \times \boldsymbol{r}_s \right\}. \tag{D.19}$$

Taking the derivative with respect to $s$ using (D.9) and (D.10), we arrive at

$$\boldsymbol{f}_\eta = -\left[ \frac{dH^2\phi}{ds} + 2H \frac{\partial H\phi}{\partial s} \right] \boldsymbol{t} + \left[ 2\frac{\partial^2 H\phi}{\partial s^2} + H^3 \phi \right] \boldsymbol{n}. \tag{D.20}$$

The above results can be used to calculate the force associated with the elastic kernel :

$$K = \kappa (H-1)^2 = \kappa \left( H^2 - H + 1 \right)^2. \tag{D.21}$$

The linear term in $H$ can be ignored because it is a topological invariant. The quadratic term can be obtained from (D.20) setting $\phi = 1$, and the constant term can be deduced from (D.11) by setting $\phi$ to zero. The result takes the form

$$f_\kappa = \kappa \left[ 2\frac{\partial^2 H}{\partial s^2} + H^3 - H \right] \boldsymbol{n} - 2\kappa \frac{dH^2}{ds} \boldsymbol{t}. \tag{D.22}$$

Finally, consider the conserved area constraint $p(A - \pi)$, where $p$ is the Lagrange multiplier and the area is given by

$$A = \frac{1}{2} \int d\theta \boldsymbol{r}^2(\theta). \tag{D.23}$$

Varying the constraint with respect to $\boldsymbol{r}$, we obtain:

$$p\delta A = p\int d\theta \boldsymbol{r} \cdot \delta \boldsymbol{r} = p\int ds \boldsymbol{n} \cdot \delta \boldsymbol{r}. \tag{D.24}$$

Hence

$$\frac{\delta p(A-\pi)}{\delta \boldsymbol{r}} = p\boldsymbol{n} \ . \tag{D.25}$$

Collecting the contributions to the forces associated with the various terms of the energy (2.4) and separating them into the normal and tangential components yields formulas (8.4) and (8.5).

## Appendix E: The Monte Carlo simulation procedure

To apply the Monte Carlo method for the model (2.4), the closed contour is discretized into $N$ sites, and the ordered sites are denoted by an index $j$ with $1 \le j \le N$. The position of each point on the contour is $\boldsymbol{r}_j = (x_j, y_j)$. At each simulation step, a site $j_0$ on the discretized contour is chosen at random. The position of the points, $\boldsymbol{r}_j$, in the vicinity of $j_0$ are displaced in a direction normal to the contour by

$$\boldsymbol{r}_j \to \boldsymbol{r}_j + \boldsymbol{n}_j \Delta r_j \qquad \text{with} \qquad \Delta r_{j_0+j} = \Delta_r \exp\left[ -\left( \frac{j_0 + j}{W} \right)^2 \right], \tag{E.1}$$

where $\boldsymbol{n}_j$ is a unit normal vector to the contour at point $j$, while the distances in the above formula are measured taking into account the periodic boundary conditions of the system. Here $\Delta_r$ is taken from a normal distribution with zero mean and variance $\sigma_r$, while the width of the normal displacement in the contour shape (9.1) is small compared to the system size, $N \gg W > 1$. This choice of a smooth update in the site positions is motivated by two reasons. First, this is the physically relevant choice, as the curvature of a real tissue can only change smoothly. Second, it is designed to ensure that the curvature of the contour changes in a controlled manner during the simulation. Since the curvature at a given point is calculated using the two nearby points, the alternative update procedure of using zero-range correlated changes in the site positions gives rise to very large curvature values unless the displacements are tiny. However, tiny displacements result in an exceedingly long time of simulation. The parameter $W$ is also used to set the short wavelength cutoff of the problem.

After the site positions are updated, the area enclosed by the contour is calculated and all site positions are rescaled using the same factor to ensure that the constant area constraint (2.5) is met.

In the next step, another site on the contour, $j_1$ is chosen randomly (independent of $j_0$), and the scalar field $\phi_j$ defined at each site is updated. In accordance with the above procedure, this field undergoes an update as follows:

$$\phi_j \to \phi_j + \Delta\phi_j \qquad \text{with} \qquad \Delta\phi_{j_1+j} = \Delta_\phi \exp\left[ -\left( \frac{j_1 + j}{W} \right)^2 \right], \tag{E.2}$$

where $\Delta_\phi$ is drawn from a normal distribution with zero mean and a variance of $\sigma_\phi$. To prevent the occurrence of negative values for the field, we repeat the selection if $\phi_j + \Delta\phi_j < 0$. The values of $\Delta_r$ and $\Delta_\phi$ are adjusted during the simulations such that the acceptance rate of the Monta Carlo changes is ½.

For the discretized versions of the unit vector normal to the contour at site $j$ we set:

$$\boldsymbol{n}_j = \frac{\boldsymbol{r}_{j+1} - \boldsymbol{r}_{j-1}}{\left|\boldsymbol{r}_{j+1} - \boldsymbol{r}_{j-1}\right|}, \tag{E.3}$$

while for the curvature, we utilize the formula

$$H_j = \frac{2}{\left|\boldsymbol{r}_j - \boldsymbol{r}_{j-1}\right|\left|\boldsymbol{r}_j - \boldsymbol{r}_{j+1}\right|\left|\boldsymbol{r}_{j+1} - \boldsymbol{r}_{j-1}\right|}\begin{vmatrix} x_{j-1} & y_{j-1} & 1 \\ x_j & y_j & 1 \\ x_{j+1} & y_{j+1} & 1 \end{vmatrix}. \tag{E.4}$$

This formula provides the inverse of the radius of a circle passing through three successive points, $\boldsymbol{r}_{j-1}$, $\boldsymbol{r}_j$ and $\boldsymbol{r}_{j+1}$. The sign of the curvature is positive if the circle lies within the domain enclosed by the contour and negative when the circle lies outside it.

The energy associated with a general form of stretching can be included by a term of the form:

$$\Theta_s = q\sum_j \left|\boldsymbol{r}_{j+1} - \boldsymbol{r}_j\right|\left(\frac{\left|\boldsymbol{r}_{j+1} - \boldsymbol{r}_j\right|}{2\pi/N} - 1\right)^{2m}, \tag{E.5}$$

where $2\pi/N$ is the initial distance between nearby points on the unit circle, while $m$ is an integer that equals one for the traditional stretching term and assumes higher values for nonlinear stretching. This term stabilizes the dynamics against the tangential forces described by Eq. (8.5). In all simulations, we have used nonlinear stretching with $q = 0.01$, $m = 2$ while the number of discretized points along the contour is $N = 300$. This choice ensures that the distance between nearby points on the contour does not become excessively large, while its influence on the contour morphological changes is negligible.